\documentclass{ectj}

\usepackage{amsfonts,amssymb,graphics,epsfig,verbatim,bm,latexsym,amsmath,url,amsbsy}

\newtheorem{theorem}{Theorem}
\newtheorem{assumption}{Assumption}
\newtheorem{proposition}{Proposition}
\newtheorem{corollary}{Corollary}
\newtheorem{lemma}{Lemma}

\newtheorem{remark}{Remark}

\renewcommand{\thesection}{\arabic{section}}
\renewcommand{\theequation}{\arabic{section}.\arabic{equation}}

\usepackage{hyperref}
\usepackage[toc,title,titletoc,header]{appendix} 
\usepackage{graphics} 

\usepackage[resetlabels]{multibib} 
\newcites{Appx}{References}

\received{May 2024}
\accepted{November 2017}
\volume{20}
  
\title[A Binary IV Model for Persuasion]{A Binary IV Model for Persuasion: Profiling Persuasion Types among Compliers}

\author[Yu]{Zeyang Yu$^{\dagger}$}

\address{$^{\dagger}$Department of Politics, Princeton University, Princeton, New Jersey, USA}
\email{arthurzeyangyu@princeton.edu}

\def\AmSTeX{$\cal A$\kern-.1667em\lower.5ex\hbox{$\cal M$}\kern-.125em
            $\cal S$-\TeX}
\def\BibTeX{{\rm B\kern-.05em{\sc i\kern-.025em b}\kern-.08em
            T\kern-.1667em\lower.7ex\hbox{E}\kern-.125emX}}

\usepackage{bbm} 
\usepackage[flushleft]{threeparttable} 
\usepackage{subcaption} 
\usepackage{graphicx} 
\usepackage[toc,title,titletoc,header]{appendix} 

\newcommand\independent{\protect\mathpalette{\protect\independenT}{\perp}}
\def\independenT#1#2{\mathrel{\rlap{$#1#2$}\mkern2mu{#1#2}}}

\DeclareMathOperator*{\cov}{Cov}
\DeclareMathOperator*{\supp}{supp}
\DeclareMathOperator*{\var}{Var}

\begin{document}

\begin{abstract}
In an empirical study of persuasion, researchers often use a binary instrument to encourage individuals to consume information and take some action. We show that, with a binary Imbens-Angrist instrumental variable model and the monotone treatment response assumption, it is possible to identify the joint distribution of potential outcomes among compliers. This is necessary to identify the percentage of mobilised voters and their statistical characteristic defined by the moments of the joint distribution of treatment and covariates. Specifically, we develop a method that enables researchers to identify the statistical characteristic of persuasion types: always-voters, never-voters, and mobilised voters among compliers. These findings extend the kappa weighting results in \cite{abadie2003semiparametric}. We also provide a sharp test for the two sets of identification assumptions. The test boils down to testing whether there exists a nonnegative solution to a possibly under-determined system of linear equations with known coefficients. An application based on \cite{green2003getting} is provided.


\end{abstract}


\section{Introduction}

Empirical studies on persuasion are widespread in economics. Researchers have examined the effect of information treatments on individual behaviour in various contexts. For example, they have studied the impact of voter mobilisation on voting, partisan media on voting, advertising on consumption, donor outreach on charity donations, and financial disclosures on investment decisions (\citealp{dellavigna2010persuasion}).

To study the problem of identifying the effects of persuasion, \cite{jun2018identifying} set up an econometric model of persuasion, which is a binary Imbens-Angrist instrumental variable (``IA IV" hereafter) model with monotone treatment response. Consider a study evaluating the effect of voter mobilisation, specifically a "Get Out the Vote" (``GOTV" hereafter) programme, on voting behaviour. In this case, the instrument is the randomly assigned information treatment, that is, mobilisation messages sent to voters. The treatment is whether voters received these messages, and the outcome is a binary behaviour measure, such as whether they voted.

This paper shows three sets of results that are economically relevant to empirical studies of persuasion. First, we show that under an econometric model of persuasion, the joint distribution of potential outcomes conditional on compliers is identified. The percentage of mobilised voters, conditional on compliers, is identifiable because the local persuasion rate is identified (\citealp{jun2018identifying}). Furthermore, under the monotone treatment response assumption, the event in which an individual is an always-voter is equivalent to the event that the individual would vote without receiving the mobilisation. A similar argument applies to never-voters.

We show that in an econometric model of persuasion, although analysts do not directly observe the three persuasion types—always-voters, never-voters, and mobilised voters among compliers—any statistical characteristic defined by the moments of the joint distribution of treatment and covariates is identifiable for these latent types. To show this, we first extend Abadie’s $\kappa$ results. Specifically, we show that in an IA IV model, any statistical characteristic defined by the moments of the joint distribution of potential outcomes, treatment, and covariates is identifiable for compliers. Building on this, we show that in an IA IV model, any statistical characteristic defined by the moments of the joint distribution of treatment and covariates can be identified, conditional on compliers and marginal potential outcomes. Finally, under the monotone treatment response assumption, we strengthen the interpretation of the conditioning set from types defined by marginal potential outcomes to types defined by joint potential outcomes.

These two new identification results are economically relevant. Persuasion involves shifting individuals from one type of action to another. To better understand the effectiveness of information treatments, researchers need insight about the joint distribution of potential outcomes (\citealp{heckman1997making}, \citealp{dellavigna2010persuasion}, \citealp{jun2018identifying}).

Additionally, profiling the statistical characteristic of persuasion types among compliers can help test the economic mechanisms through which persuasion works and is policy-relevant. Profiling these latent types can help researchers assess the mechanisms by which the treatment affects outcomes. For example, since the voter mobilisation treatment only slightly reduces voting costs, we would expect the voting propensity of mobilised voters to be marginally lower than that of always-voters. Given that voting behaviour is highly persistent, always-voters are likely to participate in multiple elections. Both hypotheses can be tested using the methods developed in this paper. These profiling results are also policy-relevant. Researchers can identify the probability of a mobilised voter being a Democrat, which allows them to estimate the number of mobilised voters who are Democrats. There are two key implications here. First, if mobilisation occurs in a swing state, the experiment could significantly influence election outcomes if mobilised voters are predominantly Democrats. Second, researchers can use these statistics to calculate the expected cost of mobilising a Democrat, enabling a better assessment of the cost-effectiveness of the intervention.

We also present three additional results that complement the analyses in \cite{jun2018identifying}. First, we show that the local persuasion rate and the approximated persuasion rate, as summarised in \cite{dellavigna2010persuasion}, are equal under one-sided non-compliance, provided that certain restrictions on the dependence between potential treatments and potential outcomes are satisfied. Next, we propose a sharp test for the identification assumptions by testing whether there are non-negative solutions to a system of linear equations implied by the model. Finally, we provide a simple sensitivity analysis for the monotone treatment response assumption.

Finally, we apply the methods to a GOTV experiment done by \cite{green2003getting}. The results indicate that approximately $10\%$ of compliers are mobilised. Among compliers, always-voters had the highest turnout in the last presidential election, while never-voters had the lowest. These findings suggest that voting behaviour is habit-forming and persistent (\citealp{gerber2003voting}). Additionally, the voting propensity of those mobilised is close to that of always-voters conditional on compliers, which is consistent with the interpretation that the GOTV programme mobilises high-propensity voters, as the intervention only slightly lowers the voting cost. In Bridgeport, the results show a high likelihood that the mobilised complier voters were Democrats, although the estimate is noisy. Based on this likelihood, we estimate that approximately 28 Democrats were mobilised due to the mobilisation, with an average cost of \$1,066 per voter.

This paper contributes to three strands of literature. First, it is closely related to \cite{abadie2003semiparametric}. We extend Abadie’s $\kappa$ results by identifying statistical characteristic measured by treatment and covariates of latent types, defined by either marginal or joint potential outcomes among compliers, under an econometric model of persuasion. This paper complements the analyses in \cite{jun2018identifying} by: (1) providing conditions under which the "approximated" persuasion rate and local persuasion rate are equal; (2) proposing a sharp test for the identification assumptions; and (3) providing a simple sensitivity analysis for the monotone treatment response assumption. Finally, this paper contributes to the literature on identifying the joint distribution of potential outcomes in an IV model. Previous work assumes rank invariance (\citealp{chernozhukov2004impact}, \citealp{vuong2017counterfactual}), while our approach restricts both the support of the outcome variable and the direction of the causal effect.

 Finally, \cite{comey2023supercompliers} independently develop methods for profiling outcome types under Assumption~\ref{assume_potential_model}, but there are four key differences between their work and ours. First, we focus on identifying the joint distribution of potential outcomes among compliers, relevant to persuasion studies, which is not explicitly stated in their paper. Second, we show that statistical characteristic defined by the moments of the joint distribution of treatment and covariates are identifiable, while they do not consider profiling with treatment. Third, we provide a sharp test for Assumption~\ref{assume_potential_model}, based on the model’s implication of an under-determined system of linear equations. The testable implication applies to an IV model under various monotonicity restrictions with a discrete outcome, a discrete instrument, and a binary treatment. Lastly, we compare different estimands relevant to persuasion studies, which they do not address.
 
The remainder of the paper is organised as follows. Section~\ref{section_model_estimands} sets up a binary IV model of persuasion. Section~\ref{sec_methodology} presents methodologies for identifying the joint distribution of potential outcomes among compliers and profiling persuasion types conditional on compliers. Additional discussions are provided in Section~\ref{sec_discussion}. In Section~\ref{sec_empirical_application}, we reanalyse the data from \cite{green2003getting} and conclude in the final section.

\section{Setup}\label{section_model_estimands}

In an empirical study of persuasion, researchers often collect data on a binary information treatment $T_{i}$, and a binary behavioural outcome $Y_{i}$. For example, in a GOTV experiment, the outcome of interest is whether or not voters vote, and the information treatment is a mobilization that contains information on the timing and the location of the upcoming election. Since information consumption involves self-selection, researchers often employ an instrument $Z_{i}$ which creates exogenous variations for an individual's information consumption decision. In many experiments, the instrument $Z_{i}$ is also binary. In a GOTV experiment, the instrument is the randomly assigned access to the GOTV treatment. Besides the aforementioned variables, researchers also collect pre-treatment covariates $X_{i} \in \mathbb{R}^{k}$. Define $Y_{i}(1)$ and $Y_{i}(0)$ as the potential outcomes that an individual would attain with and without being exposed to the treatment, and $T_{i}(1)$ and $T_{i}(0)$ as the potential treatments that an individual would attain with and without being exposed to the instrument. For a particular individual, the variable $Y_{i}(t, z)$ represents the potential outcome that this individual would obtain if $T_{i} = t$ and $Z_{i} = z$, where $t, z \in \{ 0, 1\}$.

Formally speaking, researchers make the following assumptions in a binary IV model of persuasion (\citealp{jun2018identifying}).

\begin{assumption}\label{assume_potential_model}(A Binary IV Model of Persuasion)
\begin{enumerate}
    \item Exclusion restriction: $Y_{i}(t, z) = Y_{i}(t)$, for $t, z \in \{0, 1\}$,
    \item Exogenous instrument: $Z_{i} \independent (Y_{i}(0), Y_{i}(1), T_{i}(0), T_{i}(1), X_{i})$,
    \item First stage: $\mathbb{P}[T_{i} = 1 \mid Z_{i} = 1] \neq \mathbb{P}[T_{i} = 1 \mid Z_{i} = 0]$,
    \item IV Monotonicity: $T_{i}(1) \geq T_{i}(0)$ holds almost surely,
    \item Monotone treatment response: $Y_{i}(1) \geq Y_{i}(0)$ holds almost surely, and $Y_{i}(0), Y_{i}(1) \in \{0, 1\}$.
\end{enumerate}
\end{assumption}

Assumptions 1 to 4 are the assumptions in an IA IV model. To simplify the notation, we will suppress the conditional notation throughout the paper, but all assumptions and analyses should be understood as conditional on the covariates. Note that it is not new to assume the direction of the treatment effect in causal inference literature (\citealp{manski1997monotone}, \citealp{manski2000monotone}). This assumption assumes that there are no demobilised voters.

Assumption~\ref{assume_potential_model} can be applied in contexts beyond GOTV. For instance, this model can also be used to study the persuasion effects of political messages on voting behaviour (\citealp{dellavigna2007fox}), to persuade donors to contribute (\citealp{landry2006toward}), and to assess the impact of job training programme on reducing crime (\citealp{blattman2016can}), among others.

By Assumption~\ref{assume_potential_model}, we can classify individuals into 9 groups. Since the outcome is binary, the monotone treatment response assumption implies that we can classify individuals as always-voters, never-voters, and mobilised voters. By the IV monotonicity assumption, we can classify individuals as always-takers, never-takers, and compliers. The classification is presented in Table~\ref{table_type_individual}. 

\begin{table}[ht]
\centering
\caption{Types of Individuals}
\label{table_type_individual}
\begin{tabular}{ccccll}
\multicolumn{1}{c}{$Y_{i}(0)$} & \multicolumn{1}{c}{$Y_{i}(1)$} & \multicolumn{1}{c}{$T_{i}(0)$} & \multicolumn{1}{c}{$T_{i}(1)$} & \multicolumn{1}{c}{Persuasion Type} & \multicolumn{1}{c}{Compliance Type} \\ \hline
0          & 0          & 0          & 0 & Never-Voter & Never-Taker          \\
0          & 1          & 0          & 0 & Mobilised & Never-Taker          \\
1          & 1          & 0          & 0 & Always-Voter & Never-Taker         \\
0          & 0          & 0          & 1 & Never-Voter & Complier         \\
0          & 1          & 0          & 1 & Mobilised & Complier        \\
1          & 1          & 0          & 1 & Always-Voter & Complier          \\
0          & 0          & 1          & 1 & Never-Voter & Always-Taker         \\
0          & 1          & 1          & 1 & Mobilised & Always-Taker          \\
1          & 1          & 1          & 1 & Always-Voter & Always-Taker \\ \hline
\end{tabular}
\end{table}

\section{The Proposed Methodology}\label{sec_methodology}

In this section, we present our methodology for identifying three new estimands in Section~\ref{section_model_estimands}. We first show that under Assumption~\ref{assume_potential_model}, the joint distribution of potential outcomes among compliers is identifiable. We then show how to profile persuasion types, defined by either marginal or joint potential outcomes, using pre-treatment covariates. Finally, we briefly discuss estimation and inference issues.

\subsection{Identification of the Potential Outcome Distributions for Compliers}

Remarkably, the joint distribution of potential outcomes is identified under Assumption~\ref{assume_potential_model}. In other words, under the assumptions for a binary IV model of persuasion, we can know the percentage of always-voters, never-voters, and mobilised voters among compliers. This result strengthens the classic result that identifies the quantities of the marginal distribution of the potential outcome of compliers (\citealp{imbens1994identification}, \citealp{imbens1997estimating}). We state the results formally in Proposition~\ref{corollary_local_persuasion_joint} below.

\begin{proposition}\label{corollary_local_persuasion_joint}
Under Assumption~\ref{assume_potential_model}, the joint distribution of potential outcomes among compliers is point identified:
\begin{align*}
\begin{split}
    \mathbb{P}[Y_{i}(1) = 1, Y_{i}(0) = 1 \mid T_{i}(1) > T_{i}(0)] &= \frac{\mathbb{E}[Y_{i} (1 - T_{i}) \mid Z_{i} = 0] - \mathbb{E}[Y_{i} (1 - T_{i}) \mid Z_{i} = 1]}{\mathbb{E}[T_{i} \mid Z_{i} = 1] - \mathbb{E}[T_{i} \mid Z_{i} = 0]} \\
    \mathbb{P}[Y_{i}(1) = 1, Y_{i}(0) = 0 \mid T_{i}(1) > T_{i}(0)] &= \frac{\mathbb{E}[Y_{i} \mid Z_{i} = 1] - \mathbb{E}[Y_{i} \mid Z_{i} = 0]}{\mathbb{E}[T_{i} \mid Z_{i} = 1] - \mathbb{E}[T_{i} \mid Z_{i} = 0]} \\
    \mathbb{P}[Y_{i}(1) = 0, Y_{i}(0) = 0 \mid T_{i}(1) > T_{i}(0)] &= \frac{\mathbb{E}[ (1 - Y_{i}) T_{i} \mid Z_{i} = 1] - \mathbb{E}[ (1 - Y_{i}) T_{i} \mid Z_{i} = 0]}{\mathbb{E}[T_{i} \mid Z_{i} = 1] - \mathbb{E}[T_{i} \mid Z_{i} = 0]}.
\end{split}
\end{align*}
\end{proposition}

Here is the intuition behind Proposition~\ref{corollary_local_persuasion_joint}. Under the monotone treatment response in Assumption~\ref{assume_potential_model}, those who will vote without receiving the GOTV treatment (i.e., those with $Y_{i}(0) = 1$) will also vote if they receive the GOTV treatment (i.e., their $Y_{i}(1)$ is also $1$). Therefore, identifying the proportion of always-voters among compliers boils down to identifying the proportion of voters who will vote if they do not receive the GOTV treatment among compliers, which is identifiable under the IA IV assumption (\citealp{imbens1997estimating}). Similarly, the proportion of never-voter voters among compliers is identifiable by observing that, under the monotone treatment response assumption in Assumption~\ref{assume_potential_model}, those who will not vote if they receive the GOTV treatment (i.e., those with $Y_i(1) = 0$) will also not vote if they do not receive the GOTV treatment (i.e., their $Y_i(0)$ is also $0$).

An application of Bayes' theorem shows that the proportion of persuadable individuals among compliers is the product of the local persuasion rate and the proportion of voters who will not vote without receiving the GOTV treatment among compliers. Under Assumption~\ref{assume_potential_model}, the local persuasion rate is identifiable (\citealp{jun2018identifying}). Under the IA IV assumption, the latter quantity is also identifiable (\citealp{imbens1997estimating}). Therefore, the proportion of persuadable individuals among compliers is identifiable under Assumption~\ref{assume_potential_model}.

We also discuss the extension of the identification results in Proposition~\ref{corollary_local_persuasion_joint} to non-binary outcomes and instruments in Appendix B. The results are negative for the former and positive for the latter.

\subsection{Profiling Persuasion Types}

This section presents the results that profile the persuasion types among compliers. First, we present a result that identifies any statistical characteristic of compliers, defined by the moments of the joint distribution of $(Y_{i}(t), T_{i}, X_{i})$, where $t \in \{0, 1 \}$. Then, we present a series of results identifying the statistical characteristics of compliers and persuasion types defined by the marginal potential outcomes. These statistical characteristics are measured by the moments of the joint distribution of $(T_{i}, X_{i})$. Finally, we provide results identifying the statistical characteristics of compliers and three persuasion types that are defined in Table~\ref{table_type_individual}.

\subsubsection{More on the Identification of Statistical Characteristic for Compliers}

\cite{abadie2003semiparametric} shows that any statistical characteristic that can be defined in terms of the moments of the joint distribution of $(Y_{i}(t), X_{i})$, where $t \in {0, 1}$, is identified for compliers under the IA IV assumption. We first strengthen the results in \cite{abadie2003semiparametric} by showing that any statistical characteristic that can be defined in terms of the moments of the joint distribution of $(Y_{i}(t), T_{i}, X_{i})$, where $t \in \{0, 1\}$, is identified for compliers under the IA IV assumption.

The intuition behind the results is the following. Among compliers, their treatment-taking status equals the treatment assignment. Therefore, under the IV independence assumption, identifying the moments defined by $(Y_{i}(t), T_{i}, X_{i})$ reduces to identifying the weighted averages of the moments for $(Y_{i}(t), 0, X_{i})$ and $(Y_{i}(t), 1, X_{i})$ among compliers, with the weights determined by the treatment assignment probability. Finally, \cite{abadie2003semiparametric} shows that the moments for $(Y_{i}(t), 0, X_{i})$ and $(Y_{i}(t), 1, X_{i})$ are identifiable under the IA IV assumption. We formally state the results in Theorem~\ref{thm_kappa_g_y_t_x}.

\begin{theorem}\label{thm_kappa_g_y_t_x}
Let $g(\cdot)$ be any measurable real-valued function of $(Y_{i}(t), T_{i}, X_{i})$ such that $\mathbb{E}[|g(Y_{i}(t), T_{i}, X_{i})|] < \infty$, where $t \in \{0, 1\}$. Under Assumptions 1 to 4 in Assumption~\ref{assume_potential_model}, $\mathbb{E}[g(Y_{i}(t), T_{i}, X_{i}) \mid T_{i}(1) > T_{i}(0)]$ is identifiable:
\begin{align*}
    &\mathbb{E}[g(Y_{i}(t), T_{i}, X_{i}) \mid T_{i}(1) > T_{i}(0)] \\
    &= \sum_{z \in \{0, 1\}} \left( \frac{\mathbb{E}[g(Y_{i}, z, X_{i}) \mathbbm{1}\{ T_{i} = t \} \mid Z_{i} = t ] - \mathbb{E}[g(Y_{i}, z, X_{i}) \mathbbm{1}\{ T_{i} = t \} \mid Z_{i} = 1 - t ]}{\mathbb{E}[T_{i} \mid Z_{i} = 1] - \mathbb{E}[T_{i} \mid Z_{i} = 0]} \right) \mathbb{P}[Z_{i} = z]
\end{align*}
\end{theorem}

We now present two examples that are special cases of Theorem 3.1. In the first example, we consider a function $g(\cdot)$ that is a trivial function of $T_{i}$: $g(Y_{i}(t), X_{i})$. Then, for $\mathbb{E}[g(Y_{i}(t), X_{i}) \mid T_{i}(1) > T_{i}(0)]$:
\begin{align*}
    &\mathbb{E}[g(Y_{i}(t), X_{i}) \mid T_{i}(1) > T_{i}(0)] \\
    =&\quad \frac{\mathbb{E}[g(Y_{i}, X_{i}) \mathbbm{1}\{ T_{i} = t \} \mid Z_{i} = t ] - \mathbb{E}[g(Y_{i}, X_{i}) \mathbbm{1}\{ T_{i} = t \} \mid Z_{i} = 1 - t ]}{\mathbb{E}[T_{i} \mid Z_{i} = 1] - \mathbb{E}[T_{i} \mid Z_{i} = 0]},
\end{align*}
which matches the Theorem 3.1 part (b) and part (c) in \cite{abadie2003semiparametric}. In the second example, we consider a function $g(\cdot)$ that is a trivial function of $(Y_{i}(t), X_{i})$, then, for $\mathbb{E}[T_{i} \mid T_{i}(1) > T_{i}(0)]$:
\begin{align*}
    &\mathbb{E}[T_{i} \mid T_{i}(1) > T_{i}(0)] \\
    &= \frac{\mathbb{E}[ \mathbbm{1}\{ T_{i} = t \} \mid Z_{i} = t ] - \mathbb{E}[ \mathbbm{1}\{ T_{i} = t \} \mid Z_{i} = 1 - t ]}{\mathbb{E}[T_{i} \mid Z_{i} = 1] - \mathbb{E}[T_{i} \mid Z_{i} = 0]} \mathbb{P}[Z_{i} = 1] \\
    &= \mathbb{P}[Z_{i} = 1],
\end{align*}
where the second equality uses the IV monotonicity in Assumption~\ref{assume_potential_model}.

The results in Theorem~\ref{thm_kappa_g_y_t_x} imply that we can identify any statistical characteristic defined in terms of the moments of the joint distribution of $(T_{i}, X_{i})$ for the subpopulations where $[Y_{i}(t) = y, T_{i}(1) > T_{i}(0)]$, with $t$ and $y \in \{ 0, 1 \}$. The intuition behind this result is the following. An immediate implication of Theorem~\ref{thm_kappa_g_y_t_x} is that the moments of the joint distribution of $(T_{i}, X_{i})$, conditional on compliers and a function of $Y_{i}(t)$, are also identifiable. We formally state the results in Proposition~\ref{thm_kappa_g_t_x_cond_y}.

\begin{proposition}\label{thm_kappa_g_t_x_cond_y}
Let $g(\cdot)$ be any measurable real-valued function of $(T_{i}, X_{i})$ such that $\mathbb{E}[|g(T_{i}, X_{i})|] < \infty$. Under Assumptions 1 to 4 in Assumption~\ref{assume_potential_model}, the assumption that $Y_{i}(0), Y_{i}(1) \in \{0, 1\}$, and the assumption that $\mathbb{P}[Y_{i}(t) = y, T_{i}(1) > T_{i}(0)] > 0$, where $t \text{ and } y \in \{0, 1\}$, then, $\mathbb{E}[g(T_{i}, X_{i}) \mid Y_{i}(t) = y, T_{i}(1) > T_{i}(0)]$ is identifiable:
\begin{align*}
    &\mathbb{E}[g(T_{i}, X_{i}) \mid Y_{i}(t) = y, T_{i}(1) > T_{i}(0)] \\
    &= \sum_{z \in \{0, 1\}} \left( \frac{\mathbb{E}[g(z, X_{i}) \mathbbm{1}\{Y_{i} =  y, T_{i} = t \} \mid Z_{i} = t ] - \mathbb{E}[g(z, X_{i}) \mathbbm{1}\{Y_{i} =  y, T_{i} = t \} \mid Z_{i} = 1 - t ]}{\mathbb{E}[\mathbbm{1}\{ Y_{i} = y, T_{i} = t \} \mid Z_{i} = t] - \mathbb{E}[\mathbbm{1}\{ Y_{i} = y, T_{i} = t \} \mid Z_{i} = 1- t]} \right) \mathbb{P}[Z_{i} = z].
\end{align*}
\end{proposition}

We provide some examples of $g(T_{i}, X_{i})$ below. For instance, if we choose $g(T_{i}, X_{i}) = \left(X_{i}^{j} \right)^{p}$, where $X_{i}^{j}$ is the $j$-th component of $X_{i}$ and $p \in \mathbb{R}^{+}$, we can identify any moments of the covariate $X_{i}^{j}$ if the moments exist. In a GOTV experiment, $X_{i}^{j}$ could be a binary partisanship variable, indicating whether or not $i$ is a Democrat. By choosing $p = 1$, we can identify the probability that voters belonging to the type $[Y_{i}(t) = y, T_{i}(1) > T_{i}(0)]$ are Democrats. Another example is $g(T_{i}, X_{i}) = \mathbbm{1}\{ X^{j}_{i} \leq x \}$ where $x \in \mathbb{R}$. With this choice, we can identify the cumulative distribution function of $X_{i}^{j}$ among voters belonging to the type $[Y_{i}(t) = y, T_{i}(1) > T_{i}(0)]$. For instance, if $X_{i}$ represents personal income, we can identify the cumulative density function of income among voters of the type $[Y_{i}(t) = y, T_{i}(1) > T_{i}(0)]$.

Theorem 3.1 in \cite{abadie2003semiparametric} shows that any statistical characteristic that can be defined in terms of moments of the joint distribution of $(Y_{i}, T_{i}, X_{i})$ is identified for compliers:
\begin{align*}
    \mathbb{E}[g(Y_{i}, T_{i}, X_{i}) \mid T_{i}(1) > T_{i}(0)] = \frac{1}{\mathbb{P}[T_{i}(1) > T_{i}(0)]} \mathbb{E}[\kappa g(Y_{i}, T_{i}, X_{i})],
\end{align*}
where $\kappa \equiv 1 - \frac{T_{i}(1 - Z_{i})}{\mathbb{P}[Z_{i} = 0]} - \frac{(1 - T_{i})Z_{i}}{\mathbb{P}[Z_{i} = 1]}$. Proposition~\ref{thm_kappa_g_t_x_cond_y} strengthens Abadie's $\kappa$ by further conditioning on marginal potential outcomes. Thus, a natural question is whether or not we can point identify $\mathbb{E}[g(Y_{i}, T_{i}, X_{i}) \mid Y_{i}(t) = y, T_{i}(1) > T_{i}(0)]$ under the IA IV assumption. The answer is no. To see the intuition, we use  $\mathbb{E}[g(Y_{i}, T_{i}, X_{i}) \mid Y_{i}(0) = 0, T_{i}(1) > T_{i}(0)]$ to illustrate:
\begin{align*}
\begin{split}
    &\mathbb{E}[g(Y_{i}, T_{i}, X_{i}) \mid Y_{i}(0) = 0, T_{i}(1) > T_{i}(0)] \\
    &= \mathbb{E}[g(Y_{i}(1) Z_{i} + Y_{i}(0) (1 - Z_{i}), Z_{i}, X_{i}) \mid Y_{i}(0) = 0, T_{i}(1) > T_{i}(0)] \\
    &= \mathbb{E}[g(Y_{i}(1) Z_{i}, Z_{i}, X_{i}) \mid Y_{i}(0) = 0, T_{i}(1) > T_{i}(0)] \\
    &= \mathbb{E}[g(Y_{i}(1),1, X_{i}) \mid Z_{i} = 1, Y_{i}(0) = 0, T_{i}(1) > T_{i}(0)] \mathbb{P}[Z_{i} = 1 \mid Y_{i}(0) = 0, T_{i}(1) > T_{i}(0)] \\
    &\quad + \mathbb{E}[g(0, 0, X_{i}) \mid Z_{i} = 0, Y_{i}(0) = 0, T_{i}(1) > T_{i}(0)] \mathbb{P}[Z_{i} = 0 \mid Y_{i}(0) = 0, T_{i}(1) > T_{i}(0)] \\
    &= \mathbb{E}[g(Y_{i}(1), 1, X_{i}) \mid Y_{i}(0) = 0, T_{i}(1) > T_{i}(0)] \mathbb{P}[Z_{i} = 1] \\
    &\quad + \mathbb{E}[g(0, 0, X_{i}) \mid Y_{i}(0) = 0, T_{i}(1) > T_{i}(0)] \mathbb{P}[Z_{i} = 0],
\end{split}
\end{align*}
where the first equality uses the fact that $T_{i} = Z_{i}$ for compliers, the fourth equality uses the IV independence assumption. Due to the presence of $\mathbb{E}[g(Y_{i}(1), 1, X_{i}) \mid Y_{i}(0) = 0, T_{i}(1) > T_{i}(0)] \mathbb{P}[Z_{i} = 1]$, which is about the joint distribution of potential outcomes, $\mathbb{E}[g(Y_{i}, T_{i}, X_{i}) \mid Y_{i}(0) = 0, T_{i}(1) > T_{i}(0)]$ is not point identified with the IA IV assumptions. The same intuition carries over to the remaining three cases in Proposition~\ref{thm_kappa_g_t_x_cond_y}.

Proposition~\ref{thm_kappa_g_t_x_cond_y} can be applied to continuous $Y_{i}$ by defining a new indicator variable, $\Tilde{Y}_{i} = \mathbbm{1}\{ Y_{i} \in B \}$, where $B$ is a measurable set, and a new potential outcome, $\Tilde{Y}_{i}(t) = \mathbbm{1}\{ Y_{i}(t) \in B\}$. The result in Proposition~\ref{thm_kappa_g_t_x_cond_y} holds for $\Tilde{Y}_{i}$ under the IA IV assumptions in Assumption~\ref{assume_potential_model}. An example of $B$ is: $B = \mathbbm{1}\{ Y_{i}(t) \leq \Tilde{y} \}$. That is, researchers can identify characteristics measured by $X_{i}$ of compliers and those with the potential outcome less than $\Tilde{y}$.

\subsubsection{Identification of Statistical Characteristic for Compliers and Persuasion Types}

Under Assumption~\ref{assume_potential_model}, we can identify the statistical characteristics defined by the moments of the joint distribution of $(T_{i}, X_{i})$ for always-voters, never-voters, and mobilisable voters among compliers. The intuition follows the same reasoning as in Lemma~\ref{corollary_local_persuasion_joint}. We can strengthen the interpretation of the results in Proposition~\ref{thm_kappa_g_t_x_cond_y} from conditioning on marginal potential outcomes to conditioning on joint potential outcomes under the monotone treatment response assumption in Assumption~\ref{assume_potential_model}. The results extend Theorem 3.1 in \cite{abadie2003semiparametric} by further conditioning on persuasion types defined by the pair of potential outcomes. The results are formally stated in Theorem~\ref{thm_comliance_persuasion_kappa_x_t}.

\begin{theorem}[Compliance and Persuasion]\label{thm_comliance_persuasion_kappa_x_t}
\rm Under Assumption~\ref{assume_potential_model} and the assumption that $\mathbb{P}[Y_{i}(1) = y_{1}, Y_{i}(0) = y_{0}, T_{i}(1) > T_{i}(0)] > 0$, where $y_{1} \geq y_{0}$, let $g(\cdot)$ be any measurable real-valued function of $(T_{i}, X_{i})$ such that $\mathbb{E}[|g(T_{i}, X_{i})|] < \infty$, then, the moments of $g(T_{i}, X_{i})$ conditioning on always-voters and compliers, never-voters and compliers, and mobilised compliers are identified:
\begin{align*}
    &\mathbb{E}[g(T_{i}, X_{i}) \mid Y_{i}(1) = Y_{i}(0) = 1, T_{i}(1) > T_{i}(0)] \\
    &= \sum_{z \in \{0, 1\}} \left( \frac{\mathbb{E}[g(z, X_{i}) \mathbbm{1}\{ Y_{i} = 1, T_{i} = 0 \} \mid Z_{i} = 0] - \mathbb{E}[g(z, X_{i}) \mathbbm{1}\{ Y_{i} = 1, T_{i} = 0 \} \mid Z_{i} = 1]}{\mathbb{E}[Y_{i} (1 - T_{i}) \mid Z_{i} = 0] - \mathbb{E}[Y_{i} (1 - T_{i}) \mid Z_{i} = 1]} \right) \mathbb{P}[Z_{i} = z], \\
    &\mathbb{E}[g(T_{i},X_{i}) \mid Y_{i}(1) = Y_{i}(0) = 0, T_{i}(1) > T_{i}(0)] \\
    &= \sum_{z \in \{0, 1\}} \left( \frac{\mathbb{E}[g(z, X_{i}) \mathbbm{1}\{ Y_{i} = 0, T_{i} = 1 \} \mid Z_{i} = 1] - \mathbb{E}[g(z, X_{i}) \mathbbm{1}\{ Y_{i} = 0, T_{i} = 1 \} \mid Z_{i} = 0]}{\mathbb{E}[(1 - Y_{i}) T_{i} \mid Z_{i} = 1] - \mathbb{E}[(1 - Y_{i}) T_{i} \mid Z_{i} = 0]} \right) \mathbb{P}[Z_{i} = z], \\
    &\mathbb{E}[g(T_{i},X_{i}) \mid Y_{i}(1) = 1, Y_{i}(0) = 0, T_{i}(1) > T_{i}(0)] \\
    &= \sum_{z \in \{0, 1\}} \left( \frac{\mathbb{E}[g(z, X_{i}) \mathbbm{1}\{ Y_{i} = 1 \} \mid Z_{i} = 1 ] - \mathbb{E}[g(z, X_{i}) \mathbbm{1}\{  Y_{i} = 1 \} \mid Z_{i} = 0 ]}{\mathbb{E}[Y_{i} \mid Z_{i} = 1] - \mathbb{E}[Y_{i} \mid Z_{i} = 0]} \right) \mathbb{P}[Z_{i} = z].
\end{align*}
\end{theorem}

Theorem~\ref{thm_comliance_persuasion_kappa_x_t} is a powerful identification result. Although we cannot directly observe always-voters, never-voters, or mobilised voters among compliers, we can still profile these three unobservable subpopulations using treatment and covariates. We provide three remarks on the results. First, the conditional distribution functions of a covariate given persuasion types and compliers are identified because we can define $g(T_{i}, X_{i})$ as $g(T_{i}, X_{i}) = \mathbbm{1}\{ X^{j}_{i} \leq x \}$ with $X^{j}_{i}$ is the $j$-th component of $X_{i}$ and $x \in \mathbb{R}$. Furthermore, for measurable $g$, the expectations of $g(T_{i}, X_{i})$ conditional on the three unobservable subpopulations in Theorem~\ref{thm_comliance_persuasion_kappa_x_t} are also identified, provided the expectation is well-defined. In other words, any statistical characteristics measured by the covariates $X_{i}$ of always-voters, never-voters, and mobilised voters among compliers are identified. Finally, by Bayes' rule, the conditional probability of belonging to a specific persuasion type, conditional on compliers and covariates, is also identified.

The estimands identified in Theorem~\ref{thm_comliance_persuasion_kappa_x_t} provide important insights into the intervention's impact and mechanism. For instance, in a GOTV experiment, the theorem identifies the probability that a mobilised complier is a Democrat. Although GOTV experiments are typically non-partisan, they can result in partisan outcomes, such as disproportionately mobilising Democrats. This can affect closely contested elections and helps quantify how many Democrats were mobilised, aiding analysts in evaluating the cost-effectiveness of the intervention.

Theorem~\ref{thm_comliance_persuasion_kappa_x_t} also helps assess the mechanisms by which the mobilisation affects voting. In a GOTV experiment, these results can test the hypothesis that voting is habit-forming (\citealp{gerber2003voting}). Prior voting records can serve as a measure of voting propensity, and if the hypothesis holds, always-voters among compliers should show the highest propensity, while never-voters should show the lowest.

In addition to Theorem~\ref{thm_comliance_persuasion_kappa_x_t}, there are other ways to profile voters using pre-treatment covariates. Consider this key quantity: conditional on compliers who will not vote without mobilisation, what traits define those who will vote when exposed to the treatment? Such a parameter is conditional on the voting outcome when voters are not mobilised, which is of interest to analysts focused on social justice (\citealp{heckman1997making}). For example, if a covariate measures whether a voter is African American, it allows us to determine, among compliers who would not vote without mobilisation, the percentage of additional African American voters who would vote after being mobilised. This parameter is particularly valuable for analysts seeking to better understand the effectiveness of mobilisation efforts, especially when the goal is to increase minority voter turnout. The identifiability of these estimands follows from the fact that the monotone treatment response assumption implies the identifiability of the joint distribution of the potential outcomes among compliers. These results are formally stated in Proposition~\ref{thm_profile_persudaded_types}.

\begin{proposition}\label{thm_profile_persudaded_types}
\rm Suppose Assumption~\ref{assume_potential_model} holds, let $g(\cdot)$ be any measurable real-valued function of $(T_{i}, X_{i})$ such that $\mathbb{E}[|g(T_{i}, X_{i})|] < \infty$, then, the following conditional expectations are identifiable:
\begin{align*}
    &\mathbb{E}[g(X_{i}, T_{i}) \mathbbm{1}\{ Y_{i}(1) = 0 \} \mid Y_{i}(0) = 0, T_{i}(1) > T_{i}(0) ] \\
    &= \sum_{z \in \{0, 1\}} \left( \frac{\mathbb{E}[ g(X_{i}, z) \mathbbm{1}\{ Y_{i} = 0, T_{i} = 1 \} \mid Z_{i} = 1 ] - \mathbb{E}[ g(X_{i}, z) \mathbbm{1}\{ Y_{i} = 0, T_{i} = 1 \} \mid Z_{i} = 0 ]}{\mathbb{E}[(1 - Y_{i}) (1 - T_{i}) \mid Z_{i} = 0] - \mathbb{E}[(1 - Y_{i}) (1 - T_{i}) \mid Z_{i} = 1 ]} \right) \mathbb{P}[Z_{i} = z] \\
    &\mathbb{E}[g(X_{i}, T_{i}) \mathbbm{1}\{ Y_{i}(1) = 1 \} \mid Y_{i}(0) = 0, T_{i}(1) > T_{i}(0) ] \\
    &= \sum_{z \in \{0, 1\}} \left( \frac{\mathbb{E}[g(X_{i}, z) \mathbbm{1}\{ Y_{i} = 1 \} \mid Z_{i} = 1] - \mathbb{E}[g(X_{i}, z) \mathbbm{1}\{ Y_{i} = 1 \} \mid Z_{i} = 0]}{\mathbb{E}[(1 - Y_{i}) (1 - T_{i}) \mid Z_{i} = 0] - \mathbb{E}[(1 - Y_{i}) (1 - T_{i}) \mid Z_{i} = 1]} \right) \mathbb{P}[Z_{i} = z] \\
    &\mathbb{E}[g(X_{i}, T_{i}) \mathbbm{1}\{ Y_{i}(1) = 1 \} \mid Y_{i}(0) = 1, T_{i}(1) > T_{i}(0) ] \\
    &= \sum_{z \in \{0, 1\}} \left( \frac{\mathbb{E}[g(X_{i}, z) \mathbbm{1}\{ Y_{i} = 1, T_{i} = 0 \} \mid Z_{i} = 0] - \mathbb{E}[g(X_{i}, z) \mathbbm{1}\{ Y_{i} = 1, T_{i} = 0 \} \mid Z_{i} = 1]}{\mathbb{E}[Y_{i} (1 - T_{i}) \mid Z_{i} = 0] - \mathbb{E}[Y_{i} (1 - T_{i}) \mid Z_{i} = 1]} \right) \mathbb{P}[Z_{i} = z] \\
\end{align*}
\begin{align*} 
    &\mathbb{E}[g(X_{i}, T_{i}) \mathbbm{1}\{ Y_{i}(0) = 0 \} \mid Y_{i}(1) = 0, T_{i}(1) > T_{i}(0) ] \\
    &= \sum_{z \in \{0, 1\}} \left( \frac{\mathbb{E}[g(X_{i}, z) \mathbbm{1}\{ Y_{i} = 0, T_{i} = 1 \} \mid Z_{i} = 1] - \mathbb{E}[g(X_{i}, z) \mathbbm{1}\{ Y_{i} = 0, T_{i} = 1 \} \mid Z_{i} = 0]}{\mathbb{E}[(1 - Y_{i}) T_{i} \mid Z_{i} = 1] - \mathbb{E}[(1 - Y_{i}) T_{i} \mid Z_{i} = 0]} \right) \mathbb{P}[Z_{i} = z] \\  
    &\mathbb{E}[g(X_{i}, T_{i}) \mathbbm{1}\{ Y_{i}(0) = 1 \} \mid Y_{i}(1) = 1, T_{i}(1) > T_{i}(0) ] \\
    &= \sum_{z \in \{0, 1\}} \left( \frac{\mathbb{E}[ g(X_{i}, z) \mathbbm{1}\{ Y_{i} = 1, T_{i} = 0 \} \mid Z_{i} = 0 ] - \mathbb{E}[ g(X_{i}, z) \mathbbm{1}\{ Y_{i} = 1, T_{i} = 0 \} \mid Z_{i} = 1 ]}{\mathbb{E}[Y_{i} T_{i} \mid Z_{i} = 1] - \mathbb{E}[Y_{i} T_{i} \mid Z_{i} = 0 ]} \right) \mathbb{P}[Z_{i} = z] \\ 
    &\mathbb{E}[g(X_{i}, T_{i}) \mathbbm{1}\{ Y_{i}(0) = 0 \} \mid Y_{i}(1) = 1, T_{i}(1) > T_{i}(0) ] \\
    &= \sum_{z \in \{0, 1\}} \left( \frac{\mathbb{E}[g(X_{i}, z) \mathbbm{1}\{ Y_{i} = 1 \} \mid Z_{i} = 1] - \mathbb{E}[g(X_{i}, z) \mathbbm{1}\{ Y_{i} = 1 \} \mid Z_{i} = 0]}{\mathbb{E}[Y_{i} T_{i} \mid Z_{i} = 1] - \mathbb{E}[Y_{i} T_{i} \mid Z_{i} = 0]} \right) \mathbb{P}[Z_{i} = z] 
\end{align*}
\end{proposition}

\subsection{Estimation and Inference}

This section provides estimation and inference results for the estimands we proposed. Note that the estimands we proposed in prior sections usually take the form of a Wald estimand:
\begin{align}\label{equation_wald_estimand}
    \frac{\mathbb{E}[f(X_{i}, Y_{i}, T_{i}) \mid Z_{i} = 1] - \mathbb{E}[f(X_{i}, Y_{i}, T_{i}) \mid Z_{i} = 0]}{\mathbb{E}[h(Y_{i}, T_{i}) \mid Z_{i} = 1] - \mathbb{E}[h(Y_{i}, T_{i}) \mid Z_{i} = 0]}.
\end{align}
where $f$ and $h$ are measurable functions. For example, for the case of always-voters in Theorem~\ref{thm_comliance_persuasion_kappa_x_t}, $f(X_{i}, Y_{i}, T_{i}) = \sum_{z \in \{0, 1\}} g(z, X_{i}) \mathbbm{1}\{ Y_{i} = 1, T_{i} = 0 \}$, $h(Y_{i}, T_{i}) = Y_{i} (1 - T_{i})$. It is easy to see that the numerator in Equation~\ref{equation_wald_estimand} is the coefficient of $Z_{i}$ from regressing $f(X_{i}, Y_{i}, T_{i})$ on $Z_{i}$ and a constant, while the denominator in Equation~\ref{equation_wald_estimand} is the coefficient of $Z_{i}$ from regressing $h(Y_{i}, T_{i})$ on $Z_{i}$ and a constant. Therefore, the standard estimation and inference theory for Wald estimand applies immediately to the current case with an independently and identically distributed sample of $(Y_{i}, T_{i}, Z_{i}, X_{i})$. We can either employ the conventional asymptotic results for hypothesis testing or use the Anderson-Rubin test which is robust to weak identification. We provide a more detailed discussion on inference issues in Appendix F. Note that both inferential methods can be easily implemented in standard statistical software, say, \texttt{ivreg2} and \texttt{weakiv} in \texttt{Stata}.

\section{Discussion}\label{sec_discussion}

In this section, we discuss three points on identification results from previous sections. Firstly, we compare $\theta_{\text{local}}$ with classic estimands. Additionally, we propose a test for Assumption~\ref{assume_potential_model} and a simple method to assess the sensitivity of the results to the monotone treatment response assumption.

\subsection{Comparison with Existing Estimands}

\subsubsection{Equivalence Between the Approximated Persuasion Rate and the Local Persuasion Rate Under One-Sided Non-Compliance}

As summarised in \cite{dellavigna2010persuasion}, one popular estimand in the empirics of persuasion is the ``approximated'' persuasion rate $\tilde{\theta}_{\text{DK}}$:
\begin{align*}
    \tilde{\theta}_{\text{DK}} = \frac{ \mathbb{E}[Y_{i} \mid Z_{i} = 1] - \mathbb{E}[Y_{i} \mid Z_{i} = 0] }{\mathbb{E}[T_{i} \mid Z_{i} = 1] - \mathbb{E}[T_{i} \mid Z_{i} = 0]} \times \frac{1}{1 - \mathbb{E}[Y_{i} \mid Z_{i} = 0]}.
\end{align*}
Empirical researchers often use $\tilde{\theta}_{\text{DK}}$ to approximate the persuasion rate, $\mathbb{P}[Y_{i}(1) = 1 \mid Y_{i}(0) = 0]$. However, $\theta_{\text{DK}}$ is not a well-defined conditional probability, hence, it does not measure a persuasion rate for any subpopulation (\citealp{jun2018identifying}).

Instead, \cite{jun2018identifying} propose the local persuasion rate, that measures the persuasion rate among compliers:
\begin{equation*}
    \theta_{\text{local}} \equiv \mathbb{P}[Y_{i}(1) = 1 \mid Y_{i}(0) = 0, T_{i}(1) > T_{i}(0)].
\end{equation*}
The local persuasion rate measures the percentage of compliers who take the action of interest if exposed to the treatment among those who will not take the action of interest without being exposed to the information treatment.

The results below show that under one-sided non-compliance, $\tilde{\theta}_{\text{DK}}$ equals $\theta_{\text{local}}$ under specific conditions on the distribution of potential outcomes and potential treatments. Suppose there is one-sided non-compliance in the control group. In this case, the two estimands are equal if and only if the proportion of untreated potential outcome being $0$ among untreated potential treatment being $0$ equals the proportion of never-voter among the always-takers. Suppose there is one-sided non-compliance in the treatment group. In this case, the two estimands are equivalent if and only if the untreated potential outcome is independent of the treated potential treatment.

\begin{proposition}\label{thm_equiv_theta_dk_local}
\rm Assume that Assumption~\ref{assume_potential_model} holds, if there is one-sided non-compliance in the control group, then $\theta_{\text{DK}} = \theta_{\text{local}}$ if and only if $\mathbb{P}[Y_{i}(0) = 0 \mid T_{i}(0) = 0] = \mathbb{P}[Y_{i}(1) = 0 \mid T_{i}(0) = 1]$, if there is one-sided non-compliance in the treatment group, then $\theta_{\text{DK}} = \theta_{\text{local}}$ if and only if $Y_{i}(0) \independent T_{i}(1)$.
\end{proposition}

Proposition~\ref{thm_equiv_theta_dk_local} complements the results in \cite{jun2018identifying}. \cite{jun2018identifying} show that $\tilde{\theta}_{\text{DK}} = \theta_{\text{local}}$ if certain conditions of the treatment effect homogeneity holds. Instead, after adding a one-sided non-compliance condition, Proposition~\ref{thm_equiv_theta_dk_local} shows that $\tilde{\theta}_{\text{DK}} = \theta_{\text{local}}$ if certain restrictions on the dependence between potential outcome and potential treatment hold.

\subsubsection{Complier Causal Attribution Rate}

The most closely related target parameter to the local persuasion rate is the complier causal attribution rate, which measures the proportion of observed outcome prevented by the hypothetical absence of the treatment among compliers (\citealp{yamamoto2012understanding}): 
\begin{align*}
    p_{C} = \mathbb{P}[Y_{i}(0) = 0 \mid Y_{i}(1) = 1, T_{i} = 1, T_{i}(1) > T_{i}(0)].
\end{align*}

One main difference between $p_{C}$ and $\theta_{\text{local}}$ is that the conditioning set for $p_{C}$ is $[Y_{i}(1) = 1, T_{i} = 1, T_{i} > T_{i}(0)]$ but the conditioning set for $\theta_{\text{local}}$ is $[Y_{i}(0) = 0, T_{i} > T_{i}(0)]$. Therefore, a natural way to extend the local persuasion rate is to define the local persuasion rate on the untreated:
\begin{align*}
    \theta_{\text{local untreated}} \equiv \mathbb{P}[Y_{i}(1) = 1 \mid Y_{i}(0) = 0, T_{i} = 0, T_{i}(1) > T_{i}(0)].
\end{align*}
We can point identify $\theta_{\text{local untreated}}$ given Assumption~\ref{assume_potential_model}. The intuition of the identification of $\theta_{\text{local untreated}}$ is that, conditional on compliers, $T_{i} = Z_{i}$, thus, $\theta_{\text{local untreated}} = \theta_{\text{local}}$. We formally state the result in Proposition~\ref{claim_local_untreat}.

\begin{proposition}\label{claim_local_untreat}
\rm Assume that Assumption~\ref{assume_potential_model} holds, then, $\theta_{\text{local untreated}}$ is point identifiable:
\begin{equation*}
    \theta_{\text{local untreated}} = \frac{\mathbb{E}[Y_{i} \mid Z_{i} = 1] - \mathbb{E}[Y_{i} \mid Z_{i} = 0]}{\mathbb{E}[(1 - Y_{i}) (1 - T_{i}) \mid Z_{i} = 0] - \mathbb{E}[(1 - Y_{i}) (1 - T_{i}) \mid Z_{i} = 1]}.
\end{equation*}
\end{proposition}

\subsection{A Sharp Test of the Identification Assumptions}\label{sharp_test_id_assumption}

The main identification results in Theorem~\ref{thm_comliance_persuasion_kappa_x_t} rely on two assumptions: the IA IV assumptions and the monotone treatment response assumption. These assumptions impose restrictions on individuals' choice behaviours by ruling out the defiers and the demobilised voters. Therefore, we propose a sharp test for Assumption~\ref{assume_potential_model}.

The idea of the test closely relates to \cite{balke1997bounds}. Assumptions 1, 2, 4, and 5 in Assumption~\ref{assume_potential_model} imply that the observed quantity, $\mathbb{P}[Y_{i} = y, T_{i} = t, X_{i} \in A \mid Z_{i} = z]$, with $y, t, z \in \{0, 1\}$ and $A$ measurable, is a linear combination of the probability of the unobserved persuasion and compliance types: 
\begin{align*}
    &\mathbb{P}[Y_{i} = y, T_{i} = t, X_{i} \in A \mid Z_{i} = z] \\
    &= \mathbb{P}[Y_{i}(t) = y, Y_{i}(1 - t) = y, T_{i}(z) = t, T_{i}(1 - z) = t, X_{i} \in A] \\
    &\quad + \mathbb{P}[Y_{i}(t) = y, Y_{i}(1 - t) = y, T_{i}(z) = t, T_{i}(1 - z) = 1 - t, X_{i} \in A] \\
    &\quad + \mathbb{P}[Y_{i}(t) = y, Y_{i}(1 - t) = 1 - y, T_{i}(z) = t, T_{i}(1 - z) = t, X_{i} \in A] \\
    &\quad + \mathbb{P}[Y_{i}(t) = y, Y_{i}(1 - t) = 1 - y, T_{i}(z) = t, T_{i}(1 - z) = 1 - t, X_{i} \in A].
\end{align*}
Furthermore, the defiers and the demobilised voters are ruled out by the monotonicity assumptions in Assumption~\ref{assume_potential_model} (that is, $\mathbb{P}[Y_{i}(0) = 1, Y_{i}(1) = 0] = \mathbb{P}[T_{i}(0) = 1, T_{i}(1) = 0] = 0$). Collecting these linear equations form a system of linear equations:
\begin{align*}
    A_{\text{obs}} \mathbf{p} = \mathbf{b},
\end{align*}
where $A_{\text{obs}}$ is a matrix that reflects the restrictions implied by Assumptions 1, 2, 4, and 5 in Assumption~\ref{assume_potential_model}, $\mathbf{p}$ is a non-negative vector that collects the probability of the unobserved persuasion and compliance types, and $\mathbf{b}$ is a collection of observed quantities. Note that matrix $A_{\text{obs}}$ can flexibly reflect the model restrictions. Different restrictions lead to different matrices $A_{\text{obs}}$. For example, if the model restrictions are the IA IV model without the IV relevance condition, the linear system of equations restricts the probability of defiers to zero (that is, $\mathbb{P}[T_{i}(0) = 1, T_{i}(1) = 0] = 0$).

The testable empirical implications, summarized in the system of linear equations above, provide a sharp characterization of Assumptions 1, 2, 4, and 5 in Assumption~\ref{assume_potential_model}. In other words, whenever the system of linear equations holds, there always exists another potential outcome and potential treatment model compatible with the data, in which Assumptions 1, 2, 4, and 5 of Assumption~\ref{assume_potential_model} also hold (\citealp{kitagawa2015test}, \citealp{mourifie2017testing}, \citealp{kedagni2020generalized}).

\begin{proposition}\label{prop_sharp_test}[Sharp Characterization]
(1) If Assumptions 1, 2, 4, and 5 in Assumption~\ref{assume_potential_model} hold, then, there exists $\mathbf{p} \geq \mathbf{0}$ such that $A_{\text{obs}} \mathbf{p} = \mathbf{b}$ for all measurable sets $A$. (2) For a given distribution of observables $(Y_{i}, T_{i}, X_{i}, Z_{i})$ that satisfies the restrictions in $\mathbf{P}_{0}$, then, there exists a joint distribution of $(\tilde{Y}_{i}(0), \tilde{Y}_{i}(1), \tilde{T}_{i}(0), \tilde{T}_{i}(1), X_{i}, Z_{i})$ such that Assumptions 1, 2, 4, and 5 in Assumption~\ref{assume_potential_model} hold and $(\tilde{Y}_{i}, \tilde{T}_{i}, X_{i}, Z_{i})$ has the same distribution as $(Y_{i}, T_{i}, X_{i}, Z_{i})$.
\end{proposition}

An implication of Proposition~\ref{prop_sharp_test} is that to test the validity of Assumption~\ref{assume_potential_model}, for observed data $\{ Y_{i}, T_{i}, Z_{i}, X_{i} \}_{i = 1}^{n}$ that is an independently and identically distributed sample drawn from $P \in \mathbf{P}$, it suffices to test the null hypothesis:
\begin{align}\label{iaiv_mte_h0}
    H_{0}: P \in \mathbf{P}_{0} \text{ versus } H_{1}: P \in \mathbf{P} \setminus \mathbf{P}_{0}
\end{align}
where $\mathbf{P}_{0} \equiv \{ P \in \mathbf{P}: \exists \mathbf{p} \geq \mathbf{0} \text{ s.t. } A_{\text{obs}} \mathbf{p} = \mathbf{b} \}$, which is the set of distributions that is consistent with Assumptions 1, 2, 4, and 5 in Assumption~\ref{assume_potential_model}. Thus, if $H_{0}$ is rejected, we have strong evidence against the validity of the assumptions. However, if $H_{0}$ is not rejected, we cannot confirm the validity of the assumptions. In this precise sense, Assumptions 1, 2, 4, and 5 in Assumption~\ref{assume_potential_model} are refutable but nonverifiable (\citealp{kitagawa2015test}).

In terms of the implementation of testing~\ref{iaiv_mte_h0}, with discrete $X_{i}$, we can set $A$ to be the support of $X_{i}$, and proceed the test using the recent advancement on testing whether there exists a nonnegative solution to a possibly under-determined system of linear equations with known coefficients (\citealp{bai2022testing}, \citealp{fang2023inference}). \cite{bai2022testing} propose to use subsampling method to test $H_{0}$, which can control size uniformly over $\mathbf{P}$ by the results in \cite{romano2012uniform}. The test statistic in \cite{bai2022testing} is given by:
\begin{align*}
    T_{n} \equiv \inf_{\mathbf{p} \geq \mathbf{0}: B \mathbf{p} = 1} \sqrt{n} \left| A_{\text{obs}} \mathbf{p} - \hat{\mathbf{b}} \right|,
\end{align*}
where $\hat{\mathbf{b}}$ is an estimator of $\mathbf{b}$. For more discussions on the details of computing the test statistic, see Appendix~\ref{appx_implement_sharp_test_id_assumption}. Then, consider the following quantity:
\begin{align*}
    L_{n}(t) \equiv \frac{1}{N_{n}} \sum_{1 \leq 1 \leq N_{n}} \mathbbm{1} \left\{ \inf_{\mathbf{p} \geq \mathbf{0}: B \mathbf{p} = 1} \sqrt{n} \left| A_{\text{obs}} \mathbf{p} - \hat{\mathbf{b}}_{j} \right| \leq t \right\},
\end{align*}
where $N_{n} = \binom{n}{b}$, $j$ indexes the $j$th subsample of size $b$, $\hat{\mathbf{b}}_{j}$ is $\hat{\mathbf{b}}$ evaluated at $j$th subset of the data. The subsampling-based test is:
\begin{align*}
    T_{n}^{\text{sub}} \equiv \mathbbm{1}\{ T_{n} > L_{n}^{-1}(1 - \alpha) \}.
\end{align*}

\subsection{Sensitivity Analysis: The Monotone Treatment Response Assumption}

Besides testing the identification assumptions jointly in the previous subsection, we now develop a sensitivity analysis approach to help researchers assess to what extent the point identification results are sensitive to the monotone treatment response assumption. Note that we apply the sensitivity analysis to the identification results in Lemma~\ref{corollary_local_persuasion_joint}.

The sensitivity analysis builds on the idea in \cite{balke1997bounds}. Note that the marginal distribution of the potential outcomes among compliers can be represented as the following system of linear equations:
\begin{align*}
\begin{bmatrix}
1 & 1 & 0 & 0 \\
0 & 0 & 1 & 1 \\
1 & 0 & 1 & 0 \\
0 & 1 & 0 & 1
\end{bmatrix} \begin{bmatrix}
\mathbb{P}[Y_{i}(0) = 0, Y_{i}(1) = 0 \mid T_{i}(1) > T_{i}(0)] \\
\mathbb{P}[Y_{i}(0) = 0, Y_{i}(1) = 1 \mid T_{i}(1) > T_{i}(0)] \\
\mathbb{P}[Y_{i}(0) = 1, Y_{i}(1) = 0 \mid T_{i}(1) > T_{i}(0)] \\
\mathbb{P}[Y_{i}(0) = 1, Y_{i}(1) = 1 \mid T_{i}(1) > T_{i}(0)]
\end{bmatrix} = \begin{bmatrix}
\mathbb{P}[Y_{i}(0) = 0 \mid T_{i}(1) > T_{i}(0)] \\
\mathbb{P}[Y_{i}(0) = 1 \mid T_{i}(1) > T_{i}(0)] \\
\mathbb{P}[Y_{i}(1) = 0 \mid T_{i}(1) > T_{i}(0)] \\
\mathbb{P}[Y_{i}(1) = 1 \mid T_{i}(1) > T_{i}(0)].
\end{bmatrix}
\end{align*}

Therefore, we can vary the size of $\mathbb{P}[Y_{i}(0) = 1, Y_{i}(1) = 0 \mid T_{i}(1) > T_{i}(0)]$ to see how the point identification results for the joint distribution of potential outcomes change. Here, with known $\mathbb{P}[Y_{i}(0) = 1, Y_{i}(1) = 0 \mid T_{i}(1) > T_{i}(0)]$, we can point identify $\mathbb{P}[Y_{i}(0) = 0, Y_{i}(1) = 0 \mid T_{i}(1) > T_{i}(0)]$, $\mathbb{P}[Y_{i}(0) = 0, Y_{i}(1) = 1 \mid T_{i}(1) > T_{i}(0)]$, and $\mathbb{P}[Y_{i}(0) = 1, Y_{i}(1) = 1 \mid T_{i}(1) > T_{i}(0)]$ from the system of equations above.

\section{Empirical Application: Revisit  Green et al. (2003)}\label{sec_empirical_application}

This section demonstrates the application of the methods using \cite{green2003getting} as an example. First, we provide information on the empirical setup. Then, we illustrate our main identification results with data from \cite{green2003getting}. Finally, we conduct the test for the identification assumptions and sensitivity analysis.

\subsection{Empirical Setup}

In 2001, \cite{green2003getting} conducted randomised voter mobilisation experiments during local elections in the following six cities: Bridgeport, Columbus, Detroit, Minneapolis, Raleigh, and St. Paul. Detroit, Minneapolis, and St. Paul held mayoral elections, Bridgeport had a school board election, Columbus conducted a city council election, and Raleigh hosted a mayoral/city council election. The instrument $Z_{i}$ is a randomly assigned face-to-face contact from a coalition of nonpartisan student and community organizations, encouraging voters to vote. The face-to-face contact included a brief reminder of the upcoming election in the area. The treatment $T_{i}$ is whether or not voters received the face-to-face contact. The outcome variable $Y_{i}$ is voter turnout in the local election in 2001. There are two pre-treatment covariates that we are interested in. For the full sample, we are interested in whether or not voters voted in the 2000 presidential election. We also restrict the analysis to Bridgeport. For Bridgeport, we are interested in whether or not voters are Democrats. A summary statistics table is provided in Table~\ref{tab_ggn03_sum_stats}.

\begin{table}[ht]
\centering
\begin{threeparttable}[t]
\caption{Summary Statistics in \cite{green2003getting}}
\label{tab_ggn03_sum_stats}
\begin{tabular}{lccccc} \hline
        & Observations & Mean     & Std. Dev. & Min & Max \\ \hline
\multicolumn{6}{l}{Panel A:   Full Sample}                \\ \hline
$Y_{i}$: Vote in Local Election & 18,933       & 0.296 & 0.457  & 0   & 1   \\
$T_{i}$: Take-up of the GOTV       & 18,933       & 0.135 & 0.342  & 0   & 1   \\
$Z_{i}$: Assignment of the GOTV       & 18,933       & 0.461 & 0.498   & 0   & 1   \\
Voted in 2000 & 18,933       & 0.608 & 0.488  & 0   & 1   \\ \hline
\multicolumn{6}{l}{Panel B:   Bridgeport}                 \\ \hline
$Y_{i}$: Vote in Local Election   & 1,806        & 0.118  & 0.323  & 0   & 1   \\
$T_{i}$: Take-up of the GOTV       & 1,806        & 0.137  & 0.344   & 0   & 1   \\
$Z_{i}$: Assignment of the GOTV       & 1,806        & 0.496  & 0.5  & 0   & 1   \\
Democrat     & 1,806        & 0.539 & 0.499   & 0   & 1  \\  \hline 
\end{tabular}
\begin{tablenotes}
\item Note: This table provides summary statistics for \cite{green2003getting}. Std. Dev. stands for standard deviation.
\end{tablenotes}
\end{threeparttable}
\end{table}

\subsection{Empirical Results}

We first present the results for the marginal and joint distribution of potential outcomes of compliers in Table~\ref{tab_ggn03_dist_po}. Our results reveal two interesting patterns. First, conditional on compliers, most of them are never-voters in both samples. Specifically, $61.9\%$ of voters are never-voters conditional on compliers in the full sample, and $75\%$ of voters are never-voters conditional on compliers in Bridgeport. The first stage results in Table~\ref{table-ggn03-first-stage} imply that $29.3\%$ and $27.7\%$ of voters were compliers in the full sample and the Bridgeport sample, respectively. Hence, the estimated numbers of never-voters conditional on compliers in each sample, were 3434 and 375, respectively.

\begin{table}[ht]
\centering
\begin{threeparttable}[t]
\caption{First Stage Results in \cite{green2003getting}}
\label{table-ggn03-first-stage}
\begin{tabular}{lcc} \hline
Outcome Variable: & \multicolumn{2}{c}{$T_{i}$: Take-up of the GOTV} \\ 
 & (1) & (2) \\ \hline
$Z_{i}$: Assignment of the GOTV & 0.293 & 0.277 \\
 & (0.005) & (0.015) \\
Observations & 18933 & 1806 \\
Sample & Full & Bridgeport \\ \hline
\end{tabular}
\begin{tablenotes}
\item Note: This table presents the first stage results in \cite{green2003getting}. Standard errors are presented in parentheses.
\end{tablenotes}
\end{threeparttable}
\end{table}

Second, only $7.9\%$ of voters are mobilised conditional on compliers in the full sample, and $13.9\%$ of voters are mobilised conditional on compliers in Bridgeport. Hence, the estimated number of mobilised voters conditioning on compliers in the full sample and the Bridgeport sample, were 438 and 70, respectively.

Moreover, the local persuasion rates in the full sample and the Bridgeport sample are $11.3\%$ and $15.7\%$, respectively. In other words, among the voters who are compliers and will not vote if they do not receive the GOTV intervention, $11.3\%$ and $15.7\%$ of them will vote in the full sample and the Bridgeport sample, respectively. The local persuasion rate is mechanically larger than the percentage of mobilised voters among compliers. This holds mechanically because the local persuasion rate reweights the percentage of mobilised voters by the proportion of the voters who will not vote if they do not receive the GOTV intervention.

\begin{table}[ht]
\begin{threeparttable}[t]
\caption{Distribution of Potential Outcomes in \cite{green2003getting}}
\label{tab_ggn03_dist_po}
\begin{centering}
\begin{tabular}{lccc} \hline
   & Estimates   & $95\%$ CI & $95\%$ AR CI  \\  \hline
\multicolumn{4}{l}{Panel A: Full Sample} \\  \hline
$\theta_{\text{local}}$   & 0.113       & [0.056, 0.171] &[0.054, 0.166]      \\
$\mathbb{P}[Y_{i}(0) = 1 \mid T_{i}(1) > T_{i}(0)]$   & 0.302       & [0.261, 0.343] & [0.263, 0.343]      \\
$\mathbb{P}[Y_{i}(1) = 1 \mid T_{i}(1) > T_{i}(0)]$   & 0.381       & [0.364, 0.398] &  [0.365, 0.397]     \\
$\mathbb{P}[Y_{i}(0) = 1, Y_{i}(1) = 1 \mid T_{i}(1) > T_{i}(0)]$   & 0.302       & [0.261, 0.343] & [0.263, 0.343]      \\
$\mathbb{P}[Y_{i}(0) = 0, Y_{i}(1) = 0 \mid T_{i}(1) > T_{i}(0)]$   & 0.619       & [0.602, 0.636] & [0.603, 0.635]        \\
$\mathbb{P}[Y_{i}(0) = 0, Y_{i}(1) = 1 \mid T_{i}(1) > T_{i}(0)]$   & 0.079       & [0.035, 0.123] & [0.036, 0.122]      \\  \hline
\multicolumn{4}{l}{Panel B: Bridgeport}  \\  \hline
$\theta_{\text{local}}$   & 0.157       & [0.051, 0.262] & [0.042, 0.255]      \\
$\mathbb{P}[Y_{i}(0) = 1 \mid T_{i}(1) > T_{i}(0)]$   & 0.111       & [0.019, 0.202] & [0.02, 0.202]      \\
$\mathbb{P}[Y_{i}(1) = 1 \mid T_{i}(1) > T_{i}(0)]$   & 0.25        & [0.197, 0.303] & [0.196, 0.303]      \\
$\mathbb{P}[Y_{i}(0) = 1, Y_{i}(1) = 1 \mid T_{i}(1) > T_{i}(0)]$   & 0.111       & [0.019, 0.202] & [0.02, 0.202]     \\
$\mathbb{P}[Y_{i}(0) = 0, Y_{i}(1) = 0 \mid T_{i}(1) > T_{i}(0)]$   & 0.75        & [0.697, 0.803] &  [0.697, 0.804]         \\
$\mathbb{P}[Y_{i}(0) = 0, Y_{i}(1) = 1 \mid T_{i}(1) > T_{i}(0)]$   & 0.139       & [0.033, 0.245] & [0.034, 0.244]       \\  \hline
\end{tabular}
\end{centering}
\begin{tablenotes}
\item Note: This table provides estimated marginal and joint distributions of potential outcomes among compliers for \cite{green2003getting}. CI stands for confidence interval. AR stands for Anderson-Rubin.
\end{tablenotes}
\end{threeparttable}
\end{table}

We now apply Proposition~\ref{thm_kappa_g_t_x_cond_y} and Theorem~\ref{thm_comliance_persuasion_kappa_x_t} to this experiment. The results are presented in Table~\ref{tab_ggn03_profile_x}. For the full sample, the probability of voting in the 2000 presidential election conditional on those who do not vote without the treatment and compliers  is $60.3\%$. A more interesting finding is that the subpopulation of always-voters compliers has the highest probability (that is, $95.4\%$) of voting in the 2000 presidential election. The results show that if always-voters and compliers vote in the low-profile local elections regardless of the GOTV intervention, they were very likely vote in the high-profile 2000 presidential elections. This empirical pattern is consistent with the robust findings on the persistence of voting behaviour (\citealp{gerber2003voting}). One potential explanation of the persistence of the voting behaviour is that voting behaviour is habit-forming (\citealp{gerber2003voting}). As expected, the subpopulation of never-voters and compliers has the lowest probability of voting in the 2000 presidential election.

Another interesting finding is that the voting propensity in the 2000 presidential election of the mobilised compliers is very close to the always-voters and compliers. It is consistent with the findings that GOTV experiments mobilise the high-propensity voters. One potential explanation is that the GOTV programme only mobilises the voters who are on the margin of not voting, as the intervention only slightly lowers the voting cost. Hence, the mobilised voters should have a voting propensity that is close to the always-voters.

Moreover, we also apply Abadie's $\kappa$ to identify the statistical characteristic of the compliers based on whether a voter voted in the 2000 U.S. presidential election in Table~\ref{tab_ggn03_profile_x}. The estimated likelihood that a complier voted in the 2000 U.S. presidential election is $67.4\%$. This propensity for voting is quite close to the average voting propensity in the full sample. In other words, compliers were statistically similar to the average voter in the full sample in terms of voting propensity.

In the Bridgeport sample, the most notable finding is that, among mobilised compliers, the estimated probability of being a Democrat is 81.3\%. However, the confidence interval is quite wide. Additionally, we estimate that 3.1\% of mobilised voters are also compliers and Democrats. Mobilising more Democrats in the Bridgeport school board election has practical implications for two main reasons. First, Democrats tend to be more pro-union, and second, turnout rates in these elections are typically low. For example, the turnout rate in the control group was 9.9\% (\citealp{green2003getting}). The mobilised voters might vote for pro-union candidates and help elect candidates more likely to increase teachers' salaries, benefits, and improve their working conditions (\citealp{anzia2011election}).

Beyond these benefits, the methods developed in this paper also allows researchers to assess the cost of mobilising Democrats in a GOTV experiment. Since the mobilisation message was randomly assigned with a probability of 50\%, the estimated likelihood of a Democratic, complier, and mobilised voter being mobilised is 1.6\%, or around 28 people. Assuming the experiment's costs include (1) \$3,000 for administrative expenses (including but not limited to randomisation, training canvassers, etc.), and (2) \$30 per voter for outreach, the total cost of the experiment amounts to \$29,350. Therefore, the estimated average cost of mobilising a Democrat in this study is \$1,066.

Moreover, we also utilize Abadie's $\kappa$ to identify the statistical characteristic of the compliers based on whether a voter was a Democrat in Table~\ref{tab_ggn03_profile_x}. The estimated likelihood of a complier being a Democrat is $57.3\%$. This propensity for being a Democrat is quite close to the average propensity for being a Democrat in the Bridgeport sample. In other words, compliers were statistically similar to the average voter in the Bridgeport sample in terms of the propensity for being a Democrat.

\begin{table}[ht]
\centering
\scalebox{0.83}{
\begin{threeparttable}[t]
\caption{Profiling Persuasion Types in \cite{green2003getting}}
\label{tab_ggn03_profile_x}
\begin{tabular}{lccc}  \hline
 & Estimates & $95\%$ CI & $95\%$ AR CI          \\ \hline
\multicolumn{4}{l}{Panel A: Full Sample}        \\ \hline
$\mathbb{P}[\text{Voted in 2000} = 1 \mid T_{i}(1) > T_{i}(0)]$ & 0.674 & [0.657, 0.691]  & [0.659, 0.689] \\
$\mathbb{P}[\text{Voted in 2000} = 1 \mid Y_{i}(0) = 0, T_{i}(1) > T_{i}(0)]$ & 0.603 & [0.547, 0.659]  & [0.549, 0.659] \\
$\mathbb{P}[\text{Voted in 2000} = 1 \mid Y_{i}(0) = 1, Y_{i}(1) = 1, T_{i}(1) > T_{i}(0)]$  & 0.954     & [0.914, 0.994]    & [0.914, 0.994] \\
$\mathbb{P}[\text{Voted in 2000} = 1 \mid Y_{i}(0) = 0, Y_{i}(1) = 0, T_{i}(1) > T_{i}(0)]$ & 0.511     & [0.488, 0.534]  & [0.489, 0.533] \\
$\mathbb{P}[\text{Voted in 2000} = 1 \mid Y_{i}(0) = 0, Y_{i}(1) = 1, T_{i}(1) > T_{i}(0)]$ & 0.885     & [0.715, 1]      & [0.657, 1] \\ \hline
\multicolumn{4}{l}{Panel B: Bridgeport}         \\ \hline
$\mathbb{P}[\text{Democrat} = 1 \mid T_{i}(1) > T_{i}(0)]$ & 0.573 & [0.512, 0.634]  & [0.513, 0.633] \\
$\mathbb{P}[\text{Democrat} = 1 \mid Y_{i}(0) = 0, T_{i}(1) > T_{i}(0)]$ & 0.515     & [0.35, 0.68]   & [0.349, 0.681] \\
$\mathbb{P}[\text{Democrat} = 1 \mid Y_{i}(0) = 1, Y_{i}(1) = 1, T_{i}(1) > T_{i}(0)]$  & 0.507     & [0.078, 0.935]      & [0, 0.92]     \\
$\mathbb{P}[\text{Democrat} = 1 \mid Y_{i}(0) = 0, Y_{i}(1) = 0, T_{i}(1) > T_{i}(0)]$ & 0.538     & [0.467, 0.609]   & [0.467, 0.609] \\
$\mathbb{P}[\text{Democrat} = 1 \mid Y_{i}(0) = 0, Y_{i}(1) = 1, T_{i}(1) > T_{i}(0)]$ & 0.813     & [0.437, 1]      & [0.347, 1]    \\ \hline
\end{tabular}
\begin{tablenotes}
\item Note: This table provides the results of profiling different persuasion types using pre-treatment covariates. CI refers to confidence interval. AR refers to Anderson-Rubin.
\end{tablenotes}
\end{threeparttable}}
\end{table}

\subsection{Testing Identification Assumptions and Sensitivity Analysis}

\begin{figure}[ht]

\caption{Test Identification Assumptions using \cite{bai2022testing}}

\begin{subfigure}{.49\textwidth}
  \centering
  \includegraphics[width=.9\linewidth]{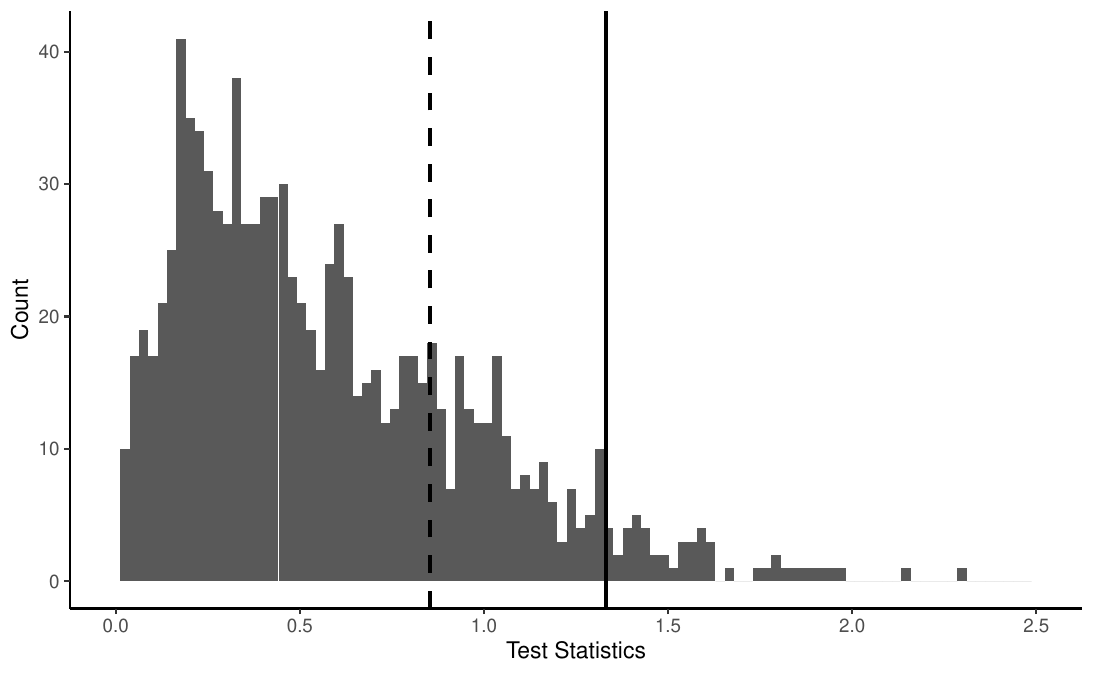}  
  \caption{Full Sample}
  \label{test_id_full_hist}
\end{subfigure}
\hfill
\begin{subfigure}{.49\textwidth}
  \centering
  \includegraphics[width=.9\linewidth]{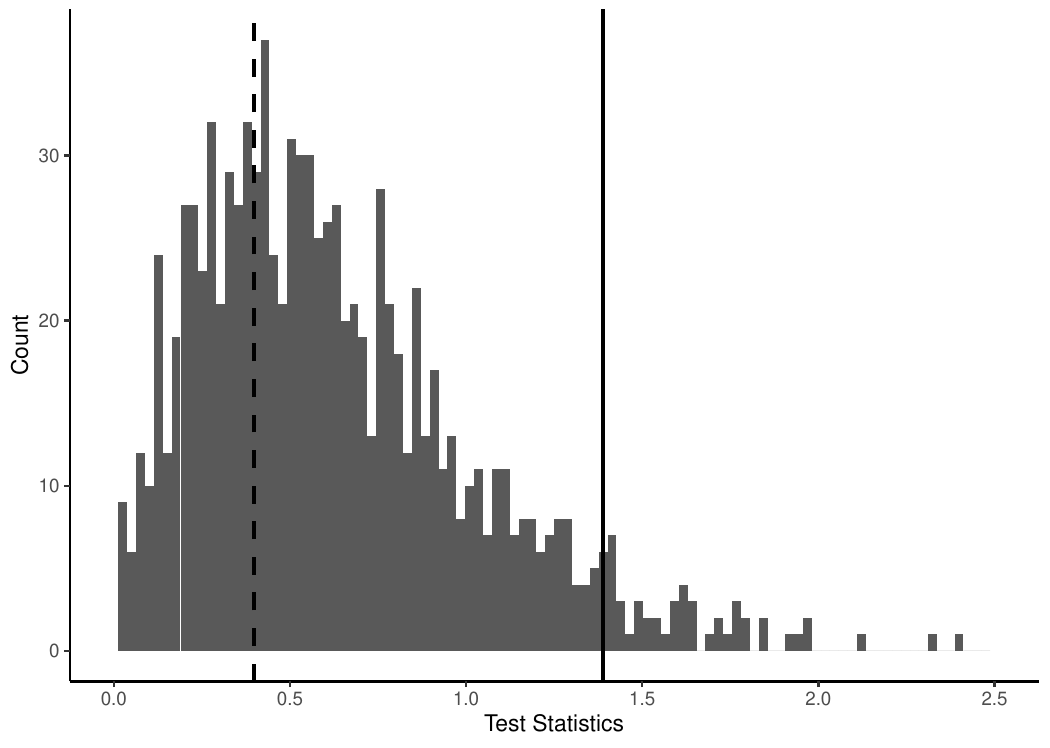}
  \caption{Bridgeport}
  \label{test_id_bridgeport_hist}
\end{subfigure}

\label{fig_test_id_assumption}
\caption*{Note. These figures present the results for testing identification assumptions. Figure~\ref{test_id_full_hist} presents the results using the full sample. Figure~\ref{test_id_bridgeport_hist} presents the results using the sample from Bridgeport. The solid lines are the critical values for a $5\%$ level test. The dashed lines are the test statistics.}
\end{figure}

We implement the test for the Assumption~\ref{assume_potential_model} by using Proposition~\ref{prop_sharp_test}. We use the subsampling method in \cite{bai2022testing} for this test. Note that the subsampling test in \cite{bai2022testing} requires us to pick a size for the subsample with $b_{n} \xrightarrow[]{} \infty$ and $\frac{b_{n}}{n} \xrightarrow[]{} 0$. We set $b_{n}$ to $n^{\frac{2}{3}}$ here. Given that the observed test statistics are less than the $95$th quantiles of the resampled distribution, results in Figure~\ref{fig_test_id_assumption} show that we cannot reject the validity of the identification assumptions at the $5\%$ level for both the full sample and the Bridgeport sample.

Furthermore, we provide the sensitivity analysis results on the joint distribution of potential outcomes in Table~\ref{tab_ggn03_sensitivity} by varying the degree to which the monotone treatment response assumption is violated among compilers. The results show that when the violation becomes larger, the proportion of mobilised voters among compliers increases.

\begin{table}[]
\begin{threeparttable}[ht]
\caption{Sensitivity for Distribution of Potential Outcomes in \cite{green2003getting}}
\label{tab_ggn03_sensitivity}
\begin{tabular}{lccccccl} \hline
\multicolumn{7}{l}{Panel A: Full   Sample}                            \\ \hline
Sensitivity Parameter &       &       &       &       &       &       \\
$\mathbb{P}[Y_{i}(1) = 0, Y_{i}(0) = 1 \mid T_{i}(1) > T_{i}(0)]$ & 0.1   & 0.12  & 0.14  & 0.16  & 0.18  & 0.2   \\ \hline
Identified Parameters &       &       &       &       &       &       \\
$\mathbb{P}[Y_{i}(1) = 1, Y_{i}(0) = 1 \mid T_{i}(1) > T_{i}(0)]$ & 0.202 & 0.182 & 0.162 & 0.142 & 0.122 & 0.102 \\
$\mathbb{P}[Y_{i}(1) = 0, Y_{i}(0) = 0 \mid T_{i}(1) > T_{i}(0)]$  & 0.519 & 0.499 & 0.479 & 0.459 & 0.439 & 0.419 \\
$\mathbb{P}[Y_{i}(1) = 1, Y_{i}(0) = 0 \mid T_{i}(1) > T_{i}(0)]$  & 0.179 & 0.199 & 0.219 & 0.239 & 0.259 & 0.279 \\ \hline
\multicolumn{7}{l}{Panel B: Bridgeport}                               \\ \hline
Sensitivity Parameter &       &       &       &       &       &       \\
$\mathbb{P}[Y_{i}(1) = 0, Y_{i}(0) = 1 \mid T_{i}(1) > T_{i}(0)]$  & 0.05  & 0.06  & 0.07  & 0.08  & 0.09  & 0.1   \\ \hline
Identified Parameters &       &       &       &       &       &       \\
$\mathbb{P}[Y_{i}(1) = 1, Y_{i}(0) = 1 \mid T_{i}(1) > T_{i}(0)]$  & 0.061 & 0.051 & 0.041 & 0.031 & 0.021 & 0.011 \\
$\mathbb{P}[Y_{i}(1) = 0, Y_{i}(0) = 0 \mid T_{i}(1) > T_{i}(0)]$  & 0.7   & 0.69  & 0.68  & 0.67  & 0.66  & 0.65  \\
$\mathbb{P}[Y_{i}(1) = 1, Y_{i}(0) = 0 \mid T_{i}(1) > T_{i}(0)]$  & 0.189 & 0.199 & 0.209 & 0.219 & 0.229 & 0.239 \\  \hline
\end{tabular}
\begin{tablenotes}
\item Note: This table provides sensitivity analysis on the joint distribution of potential outcomes among compliers by varying the size of the dissuaded among compliers.
\end{tablenotes}
\end{threeparttable}
\end{table}

\section{Concluding Remarks}

This paper studies a binary IA IV model for persuasion. We show that in an econometric model of persuasion, it is possible to identify the joint distribution of potential outcomes among compliers. We develop a weighting method that helps researchers identify the statistical characteristics of persuasion types: always-voters and compliers, never-voters and compliers, and mobilised compliers. We apply the proposed methodology to profile persuasion types in the GOTV experiments by \cite{green2003getting}. The results show that among compliers, roughly $10\%$ voters are mobilised. The results are also consistent with the findings that voters' voting behaviours are highly persistent and the mobilised voters are high propensity voters. In future research, researchers can apply the partial identification approach proposed by \cite{mogstad2018using} to assess the welfare impact of the information treatment by partially identifying the persuasion rate, either with or without assuming a monotone treatment response.

\section*{Acknowledgements}

I thank my advisors, Scott Gehlbach, Robert Gulotty, and Alexander Torgovitsky, who were gracious with their advice, support, and feedback. I additionally thank Alberto Abadie, Eric Auerbach, Stephane Bonhomme, Federico Bugni, Joshua Byun, Matias Cattaneo, Gustavo Diaz, Yingying Dong, Wayne Yuan Gao, Justin Grimmer, Peter Hull, Kosuke Imai, Sung Jae Jun, Michal Kolesar, Nadav Kunievsky, Xinran Li, Jonathan Mummolo,  Molly Offer-Westort, Zhuan Pei, Maggie Penn, Kirill Ponomarev, James Robinson, Jonathan Roth, Francesco Ruggieri, Azeem Shaikh, Tymon Sloczynski, Joshua Ka Chun Shea, Liyang Sun, Max Tabord-Meehan, Christopher Walters, Linbo Wang, Yiqing Xu, Teppei Yamamoto, Boyang Zhang, as well as participants in conferences and seminars at the American Causal Inference Conference 2022, the Midwest Econometrics Group Conference 2022, Harvard, and Princeton, for their helpful comments on this paper. Finally, I thank the co-editor and the two anonymous referees for their constructive comments, which have substantially improved the paper.

\bibliography{reference}       

\begin{thebibliography}{}

\bibitem[\protect\citeauthoryear{Abadie}{Abadie}{2003}]{abadie2003semiparametric}
Abadie, A. (2003).
\newblock Semiparametric instrumental variable estimation of treatment response models.
\newblock {\em Journal of Econometrics\/}~{\em 113\/}(2), 231--263.

\bibitem[\protect\citeauthoryear{Anzia}{Anzia}{2011}]{anzia2011election}
Anzia, S.~F. (2011).
\newblock Election timing and the electoral influence of interest groups.
\newblock {\em Journal of Politics\/}~{\em 73\/}(2), 412--427.

\bibitem[\protect\citeauthoryear{Bai, Santos, and Shaikh}{Bai et~al.}{2022}]{bai2022testing}
Bai, Y., A.~Santos, and A.~M. Shaikh (2022).
\newblock On testing systems of linear inequalities with known coefficients.
\newblock {\em Working Paper\/}.

\bibitem[\protect\citeauthoryear{Balke and Pearl}{Balke and Pearl}{1997}]{balke1997bounds}
Balke, A. and J.~Pearl (1997).
\newblock Bounds on treatment effects from studies with imperfect compliance.
\newblock {\em Journal of the American Statistical Association\/}~{\em 92\/}(439), 1171--1176.

\bibitem[\protect\citeauthoryear{Blattman and Annan}{Blattman and Annan}{2016}]{blattman2016can}
Blattman, C. and J.~Annan (2016).
\newblock Can employment reduce lawlessness and rebellion? a field experiment with high-risk men in a fragile state.
\newblock {\em American Political Science Review\/}~{\em 110\/}(1), 1--17.

\bibitem[\protect\citeauthoryear{Chernozhukov and Hansen}{Chernozhukov and Hansen}{2004}]{chernozhukov2004impact}
Chernozhukov, V. and C.~Hansen (2004).
\newblock The impact of 401 (k) participation on the wealth distribution: An instrumental quantile regression analysis.
\newblock {\em Review of Economics and Statistics\/}~{\em 86\/}(3), 735--751.

\bibitem[\protect\citeauthoryear{Comey, Eng, and Pei}{Comey et~al.}{2023}]{comey2023supercompliers}
Comey, M.~L., A.~R. Eng, and Z.~Pei (2023).
\newblock Supercompliers.
\newblock {\em arXiv preprint arXiv:2212.14105\/}.

\bibitem[\protect\citeauthoryear{DellaVigna and Gentzkow}{DellaVigna and Gentzkow}{2010}]{dellavigna2010persuasion}
DellaVigna, S. and M.~Gentzkow (2010).
\newblock Persuasion: empirical evidence.
\newblock {\em Annual Review of Economics\/}~{\em 2\/}(1), 643--669.

\bibitem[\protect\citeauthoryear{DellaVigna and Kaplan}{DellaVigna and Kaplan}{2007}]{dellavigna2007fox}
DellaVigna, S. and E.~Kaplan (2007).
\newblock The fox news effect: Media bias and voting.
\newblock {\em The Quarterly Journal of Economics\/}~{\em 122\/}(3), 1187--1234.

\bibitem[\protect\citeauthoryear{Fang, Santos, Shaikh, and Torgovitsky}{Fang et~al.}{2023}]{fang2023inference}
Fang, Z., A.~Santos, A.~M. Shaikh, and A.~Torgovitsky (2023).
\newblock Inference for large-scale linear systems with known coefficients.
\newblock {\em Econometrica\/}~{\em 91\/}(1), 299--327.

\bibitem[\protect\citeauthoryear{Gerber, Green, and Shachar}{Gerber et~al.}{2003}]{gerber2003voting}
Gerber, A.~S., D.~P. Green, and R.~Shachar (2003).
\newblock Voting may be habit-forming: evidence from a randomized field experiment.
\newblock {\em American Journal of Political Science\/}~{\em 47\/}(3), 540--550.

\bibitem[\protect\citeauthoryear{Green, Gerber, and Nickerson}{Green et~al.}{2003}]{green2003getting}
Green, D.~P., A.~S. Gerber, and D.~W. Nickerson (2003).
\newblock Getting out the vote in local elections: Results from six door-to-door canvassing experiments.
\newblock {\em Journal of Politics\/}~{\em 65\/}(4), 1083--1096.

\bibitem[\protect\citeauthoryear{Heckman, Smith, and Clements}{Heckman et~al.}{1997}]{heckman1997making}
Heckman, J.~J., J.~Smith, and N.~Clements (1997).
\newblock Making the most out of programme evaluations and social experiments: Accounting for heterogeneity in programme impacts.
\newblock {\em Review of Economic Studies\/}~{\em 64\/}(4), 487--535.

\bibitem[\protect\citeauthoryear{Imbens and Angrist}{Imbens and Angrist}{1994}]{imbens1994identification}
Imbens, G.~W. and J.~D. Angrist (1994).
\newblock Identification and estimation of local average treatment effects.
\newblock {\em Econometrica\/}, 467--475.

\bibitem[\protect\citeauthoryear{Imbens and Rubin}{Imbens and Rubin}{1997}]{imbens1997estimating}
Imbens, G.~W. and D.~B. Rubin (1997).
\newblock Estimating outcome distributions for compliers in instrumental variables models.
\newblock {\em Review of Economic Studies\/}~{\em 64\/}(4), 555--574.

\bibitem[\protect\citeauthoryear{Jun and Lee}{Jun and Lee}{2023}]{jun2018identifying}
Jun, S.~J. and S.~Lee (2023).
\newblock Identifying the effect of persuasion.
\newblock {\em Journal of Political Economy\/}~{\em 131\/}(8), 2032--2058.

\bibitem[\protect\citeauthoryear{K{\'e}dagni and Mourifi{\'e}}{K{\'e}dagni and Mourifi{\'e}}{2020}]{kedagni2020generalized}
K{\'e}dagni, D. and I.~Mourifi{\'e} (2020).
\newblock Generalized instrumental inequalities: testing the instrumental variable independence assumption.
\newblock {\em Biometrika\/}~{\em 107\/}(3), 661--675.

\bibitem[\protect\citeauthoryear{Kitagawa}{Kitagawa}{2015}]{kitagawa2015test}
Kitagawa, T. (2015).
\newblock A test for instrument validity.
\newblock {\em Econometrica\/}~{\em 83\/}(5), 2043--2063.

\bibitem[\protect\citeauthoryear{Landry, Lange, List, Price, and Rupp}{Landry et~al.}{2006}]{landry2006toward}
Landry, C.~E., A.~Lange, J.~A. List, M.~K. Price, and N.~G. Rupp (2006).
\newblock Toward an understanding of the economics of charity: Evidence from a field experiment.
\newblock {\em Quarterly Journal of Economics\/}~{\em 121\/}(2), 747--782.

\bibitem[\protect\citeauthoryear{Manski}{Manski}{1997}]{manski1997monotone}
Manski, C. (1997).
\newblock Monotone treatment response.
\newblock {\em Econometrica\/}~{\em 65\/}(6), 1311--1334.

\bibitem[\protect\citeauthoryear{Manski and Pepper}{Manski and Pepper}{2000}]{manski2000monotone}
Manski, C.~F. and J.~V. Pepper (2000).
\newblock Monotone instrumental variables: With an application to the returns to schooling.
\newblock {\em Econometrica\/}~{\em 68\/}(4), 997--1010.

\bibitem[\protect\citeauthoryear{Mogstad, Santos, and Torgovitsky}{Mogstad et~al.}{2018}]{mogstad2018using}
Mogstad, M., A.~Santos, and A.~Torgovitsky (2018).
\newblock Using instrumental variables for inference about policy relevant treatment parameters.
\newblock {\em Econometrica\/}~{\em 86\/}(5), 1589--1619.

\bibitem[\protect\citeauthoryear{Mourifi{\'e} and Wan}{Mourifi{\'e} and Wan}{2017}]{mourifie2017testing}
Mourifi{\'e}, I. and Y.~Wan (2017).
\newblock Testing local average treatment effect assumptions.
\newblock {\em Review of Economics and Statistics\/}~{\em 99\/}(2), 305--313.

\bibitem[\protect\citeauthoryear{Romano and Shaikh}{Romano and Shaikh}{2012}]{romano2012uniform}
Romano, J.~P. and A.~M. Shaikh (2012).
\newblock On the uniform asymptotic validity of subsampling and the bootstrap.
\newblock {\em Annals of Statistics\/}~{\em 40\/}(6), 2798--2822.

\bibitem[\protect\citeauthoryear{Vuong and Xu}{Vuong and Xu}{2017}]{vuong2017counterfactual}
Vuong, Q. and H.~Xu (2017).
\newblock Counterfactual mapping and individual treatment effects in nonseparable models with binary endogeneity.
\newblock {\em Quantitative Economics\/}~{\em 8\/}(2), 589--610.

\bibitem[\protect\citeauthoryear{Yamamoto}{Yamamoto}{2012}]{yamamoto2012understanding}
Yamamoto, T. (2012).
\newblock Understanding the past: Statistical analysis of causal attribution.
\newblock {\em American Journal of Political Science\/}~{\em 56\/}(1), 237--256.

\end{thebibliography}


\begin{thebibliography}{}

\bibitem[\protect\citeauthoryear{Abadie}{Abadie}{2003}]{abadie2003semiparametric}
Abadie, A. (2003).
\newblock Semiparametric instrumental variable estimation of treatment response models.
\newblock {\em Journal of Econometrics\/}~{\em 113\/}(2), 231--263.

\bibitem[\protect\citeauthoryear{Balke and Pearl}{Balke and Pearl}{1997}]{balke1997bounds}
Balke, A. and J.~Pearl (1997).
\newblock Bounds on treatment effects from studies with imperfect compliance.
\newblock {\em Journal of the American Statistical Association\/}~{\em 92\/}(439), 1171--1176.

\bibitem[\protect\citeauthoryear{Boyd and Vandenberghe}{Boyd and Vandenberghe}{2004}]{boyd2004convex}
Boyd, S. and L.~Vandenberghe (2004).
\newblock {\em Convex Optimization}.
\newblock Cambridge: Cambridge university press.

\bibitem[\protect\citeauthoryear{Carneiro and Lee}{Carneiro and Lee}{2009}]{carneiro2009estimating}
Carneiro, P. and S.~Lee (2009).
\newblock Estimating distributions of potential outcomes using local instrumental variables with an application to changes in college enrollment and wage inequality.
\newblock {\em Journal of Econometrics\/}~{\em 149\/}(2), 191--208.

\bibitem[\protect\citeauthoryear{Durrett}{Durrett}{2010}]{durrett2010probability}
Durrett, R. (2010).
\newblock {\em Probability: Theory and Examples}.
\newblock Cambridge: Cambridge university press.

\bibitem[\protect\citeauthoryear{Fu, Narasimhan, and Boyd}{Fu et~al.}{2020}]{fu2017cvxr}
Fu, A., B.~Narasimhan, and S.~Boyd (2020).
\newblock Cvxr: An r package for disciplined convex optimization.
\newblock {\em Journal of Statistical Software\/}~{\em 94}, 1--34.

\bibitem[\protect\citeauthoryear{Hansen}{Hansen}{2022}]{hansen2022econometrics}
Hansen, B. (2022).
\newblock {\em Econometrics}.
\newblock Princeton University Press.

\bibitem[\protect\citeauthoryear{Heckman and Vytlacil}{Heckman and Vytlacil}{2005}]{heckman2005structural}
Heckman, J.~J. and E.~Vytlacil (2005).
\newblock Structural equations, treatment effects, and econometric policy evaluation 1.
\newblock {\em Econometrica\/}~{\em 73\/}(3), 669--738.

\bibitem[\protect\citeauthoryear{Imbens and Rubin}{Imbens and Rubin}{1997}]{imbens1997estimating}
Imbens, G.~W. and D.~B. Rubin (1997).
\newblock Estimating outcome distributions for compliers in instrumental variables models.
\newblock {\em The Review of Economic Studies\/}~{\em 64\/}(4), 555--574.

\bibitem[\protect\citeauthoryear{Jun and Lee}{Jun and Lee}{2023}]{jun2018identifying}
Jun, S.~J. and S.~Lee (2023).
\newblock Identifying the effect of persuasion.
\newblock {\em Journal of Political Economy\/}~{\em 131\/}(8), 2032--2058.

\bibitem[\protect\citeauthoryear{Manski}{Manski}{1997}]{manski1997monotone}
Manski, C. (1997).
\newblock Monotone treatment response.
\newblock {\em Econometrica\/}~{\em 65\/}(6), 1311--1334.

\bibitem[\protect\citeauthoryear{Staiger and Stock}{Staiger and Stock}{1997}]{staiger1997instrumental}
Staiger, D. and J.~H. Stock (1997).
\newblock Instrumental variables regression with weak instruments.
\newblock {\em Econometrica\/}~{\em 65\/}(3), 557--586.

\end{thebibliography}


\newpage

\begin{appendices}

\setcounter{page}{1}

\section*{Online Appendix For "A Binary IV Model for Persuasion"}

Appendix~\ref{appx_proof} contains the proofs for the main results in the paper as well as additional results referenced or provided in the appendix. Appendix~\ref{appx_identify_nonbinary_z_y} addresses the identifiability of the joint distribution of potential outcomes with non-binary instruments or outcomes. Appendix~\ref{appx_profiling_nonbinary_z} discusses the results on profiling persuasion types conditional on compliers using covariates. Appendix~\ref{id_at_nt} discusses the results on profiling persuasion types conditional on always-takers and never-takers. Appendix~\ref{appx_estimation_inference} provides further details on estimation and inference. Finally, Appendix~\ref{appx_implement_sharp_test_id_assumption} provides computational details for implementing the tests for the identification assumptions.

\section{Proofs}\label{appx_proof}

\subsection{Lemma~\ref{corollary_marginal_potential} and Its Proof}\label{appx_pf_complier_marginal_po}

Proposition~\ref{corollary_local_persuasion_joint} uses a classic result: under the IA and IA IV assumptions, the marginal distribution of potential outcomes among compliers can be point identified (\citealpAppx{abadie2003semiparametric}, \citealpAppx{imbens1997estimating}). We now state and prove this classic result for completeness.

\begin{lemma}\label{corollary_marginal_potential}
Assume that the 1 to 4 in Assumption~\ref{assume_potential_model} hold, the marginal distribution of potential outcomes conditioning on compliers is point identified:
\begin{align*}
    \mathbb{P}[Y_{i}(0) = y \mid T_{i}(1) > T_{i}(0)] &= \frac{\mathbb{E}[ \mathbbm{1}\{ Y_{i} = y \} (1 - T_{i}) \mid Z_{i} = 0] - \mathbb{E}[ \mathbbm{1}\{ Y_{i} = y \} (1 - T_{i}) \mid Z_{i} = 1]}{\mathbb{E}[T_{i} \mid Z_{i} = 1] - \mathbb{E}[T_{i} \mid Z_{i} = 0]} \\
    \mathbb{P}[Y_{i}(1) = y \mid T_{i}(1) > T_{i}(0)] &= \frac{\mathbb{E}[ \mathbbm{1}\{ Y_{i} = y \} T_{i} \mid Z_{i} = 1] - \mathbb{E}[ \mathbbm{1}\{ Y_{i} = y \} T_{i} \mid Z_{i} = 0]}{\mathbb{E}[T_{i} \mid Z_{i} = 1] - \mathbb{E}[T_{i} \mid Z_{i} = 0]},
\end{align*}
where $y \in \{0, 1\}$.
\end{lemma}

\begin{proof}
For $\mathbb{P}[Y_{i}(t) = y | T_{i}(1) > T_{i}(0)]$ where $y \in \{0, 1\}$ and $t \in \{0, 1\}$, we have the following:
\begin{align*}
    \mathbb{P}[Y_{i}(t) = y | T_{i}(1) > T_{i}(0)] &= \frac{\mathbb{P}[Y_{i}(t) = y, T_{i}(1) = 1, T_{i}(0) = 0]}{\mathbb{P}[T_{i}(1) = 1, T_{i}(0) = 0]} \\
    &= \frac{\mathbb{P}[Y_{i}(t) = y, T_{i}(1) = 1, T_{i}(0) = 0]}{\mathbb{E}[T_{i} | Z_{i} = 1] - \mathbb{E}[T_{i} | Z_{i} = 0]},
\end{align*}
where the second equality uses Lemma 2.1 in \citeAppx{abadie2003semiparametric}.

For $\mathbb{P}[Y_{i}(t) = y, T_{i}(1) = 1, T_{i}(0) = 0]$ with $y \in \{0, 1\}$ and $t \in \{0, 1\}$:
\begin{align*}
    &\mathbb{P}[Y_{i}(t) = y, T_{i}(1) = 1, T_{i}(0) = 0] \\
    &= \mathbb{P}[Y_{i}(t) = y, T_{i}(t) = t] - \mathbb{P}[Y_{i}(t) = y, T_{i}(t) = t, T_{i}(1 - t) = t] \\
    &= \mathbb{P}[Y_{i}(t) = y, T_{i}(t) = t] - \mathbb{P}[Y_{i}(t) = y, T_{i}(1 - t) = t] \\
    &= \mathbb{P}[Y_{i}(t) = y, T_{i}(t) = t | Z_{i} = t] - \mathbb{P}[Y_{i}(t) = y, T_{i}(1 - t) = t | Z_{i} = 1 - t] \\
    &= \mathbb{P}[Y_{i} = y, T_{i} = t | Z_{i} = t] - \mathbb{P}[Y_{i} = y, T_{i} = t | Z_{i} = 1 - t],
\end{align*}
where the first and the second equality uses IV monotonicity in Assumption~\ref{assume_potential_model}, the third equality uses IV exogeneity in Assumption~\ref{assume_potential_model}. Now, the results follow immediately. \qed
\end{proof}

\subsection{Proof of Lemma~\ref{corollary_local_persuasion_joint}}

The monotone treatment response assumption in Assumption~\ref{assume_potential_model} implies that $[Y_{i}(1) = 1, Y_{i}(0) = 1] = [Y_{i}(0) = 1]$. Therefore, to identify $\mathbb{P}[Y_{i}(1) = 1, Y_{i}(0) = 1 \mid T_{i}(1) > T_{i}(0)]$, it suffices to identify $\mathbb{P}[Y_{i}(0) = 1 \mid T_{i}(1) > T_{i}(0)]$, which is identifiable by Lemma~\ref{corollary_marginal_potential}.

The result for $\mathbb{P}[Y_{i}(1) = 0, Y_{i}(0) = 0 \mid T_{i}(1) > T_{i}(0)]$ can be derived analogously by observing that monotone treatment response assumption in Assumption~\ref{assume_potential_model} implies $[Y_{i}(1) = 0, Y_{i}(0) = 0] = [Y_{i}(1) = 0]$ and using Lemma~\ref{corollary_marginal_potential}.

For $\mathbb{P}[Y_{i}(1) = 1, Y_{i}(0) = 0 \mid T_{i}(1) > T_{i}(0)]$:
\begin{align*}
    &\mathbb{P}[Y_{i}(1) = 1, Y_{i}(0) = 0 \mid T_{i}(1) > T_{i}(0)] \\
    &= \mathbb{P}[Y_{i}(1) = 1  \mid Y_{i}(0) = 0, T_{i}(1) > T_{i}(0)] \times \mathbb{P}[Y_{i}(0) = 0 \mid T_{i}(1) > T_{i}(0)] \\
    &= \frac{\mathbb{E}[Y_{i} \mid Z_{i} = 1] - \mathbb{E}[Y_{i} \mid Z_{i} = 0]}{\mathbb{E}[ \mathbbm{1}\{ Y_{i} = 0 \} (1 - T_{i}) \mid Z_{i} = 0] - \mathbb{E}[ \mathbbm{1}\{ Y_{i} = 0 \} (1 - T_{i}) \mid Z_{i} = 1]} \\
    &\quad \times \frac{\mathbb{E}[ \mathbbm{1}\{ Y_{i} = 0 \} (1 - T_{i}) \mid Z_{i} = 0] - \mathbb{E}[ \mathbbm{1}\{ Y_{i} = 0 \} (1 - T_{i}) \mid Z_{i} = 1]}{\mathbb{E}[T_{i} \mid Z_{i} = 1] - \mathbb{E}[T_{i} \mid Z_{i} = 0]} \\
    &= \frac{\mathbb{E}[Y_{i} \mid Z_{i} = 1] - \mathbb{E}[Y_{i} \mid Z_{i} = 0]}{\mathbb{E}[T_{i} \mid Z_{i} = 1] - \mathbb{E}[T_{i} \mid Z_{i} = 0]},
\end{align*}
where the second equality uses Theorem 6 in \citeAppx{jun2018identifying} and Lemma~\ref{corollary_marginal_potential}. \qed

\begin{remark}
As suggested by one of the referees, there are alternative ways to prove Lemma~\ref{corollary_local_persuasion_joint}. First, the monotone treatment response assumption in Assumption~\ref{assume_potential_model} implies that $\mathbb{P}[Y_{i}(1) = 1, Y_{i}(0) = 0 \mid T_{i}(1) > T_{i}(0)] = \mathbb{E}[Y_{i}(1) - Y_{i}(0) \mid T_{i}(1) > T_{i}(0)]$, and this latter quantity, which is the local average treatment effect, is identifiable under an IA IV model. Another way to show the result is by observing that the monotone treatment response assumption in Assumption~\ref{assume_potential_model} implies $\mathbb{P}[Y_{i}(1) = 0, Y_{i}(0) = 1 \mid T_{i}(1) > T_{i}(0)] = 0$. Therefore, $\mathbb{P}[Y_{i}(1) = 1, Y_{i}(0) = 0 \mid T_{i}(1) > T_{i}(0)] = 1 - \mathbb{P}[Y_{i}(1) = 1, Y_{i}(0) = 1 \mid T_{i}(1) > T_{i}(0)] - \mathbb{P}[Y_{i}(1) = 0, Y_{i}(0) = 0 \mid T_{i}(1) > T_{i}(0)]$, where the latter two probabilities are identifiable under Assumption~\ref{assume_potential_model}.
\end{remark}

\subsection{Proof of Theorem~\ref{thm_kappa_g_y_t_x}}

Observe that by Lemma 3.1 in \cite{abadie2003semiparametric}:
\begin{align*}
    \mathbb{E}[g(Y_{i}(t), T_{i}, X_{i}) \mid T_{i}(1) > T_{i}(0)] = \frac{\mathbb{E}[g(Y_{i}(t), T_{i}, X_{i}) \mathbbm{1}\{ T_{i}(1) > T_{i}(0) \}]}{\mathbb{E}[T_{i} \mid Z_{i} = 1] - \mathbb{E}[T_{i} \mid Z_{i} = 0]}.
\end{align*}

For $\mathbb{E}[g(Y_{i}(t), T_{i}, X_{i}) \mathbbm{1}\{ T_{i}(1) > T_{i}(0) \}]$:
\begin{align*}
    &\mathbb{E}[g(Y_{i}(t), T_{i}, X_{i}) \mathbbm{1}\{ T_{i}(1) > T_{i}(0) \}] \\
    &= \sum_{z \in \{ 0, 1 \} } \mathbb{E}[g(Y_{i}(t), z, X_{i}) \mathbbm{1}\{ T_{i}(1) > T_{i}(0) \} \mathbbm{1}\{ Z_{i} = z \} ] \\
    &= \sum_{z \in \{ 0, 1 \} } \mathbb{E}[g(Y_{i}(t), z, X_{i}) \mathbbm{1}\{ T_{i}(1) > T_{i}(0) \}] \mathbb{P}[Z_{i} = z],
\end{align*}
where the first equality uses the fact that $T_{i} = Z_{i}$ among compliers, and the second equality uses the IV exogeneity assumption.

For $\mathbb{E}[g(Y_{i}(t), z, X_{i}) \mathbbm{1}\{ T_{i}(1) > T_{i}(0) \} ]$, where $t \in \{0, 1 \}$ and $z \in \{0, 1\}$:
\begin{align*}
    &\mathbb{E}[g(Y_{i}(t), z, X_{i}) \mathbbm{1}\{ T_{i}(1) > T_{i}(0) \} ] \\
    &= \mathbb{E}[g(Y_{i}(t), z, X_{i}) (\mathbbm{1}\{ T_{i}(t) = t \} - \mathbbm{1}\{ T_{i}(t) = t, T_{i}(1 - t) = t \})] \\
    &= \mathbb{E}[g(Y_{i}(t), z, X_{i}) (\mathbbm{1}\{ T_{i}(t) = t \} - \mathbbm{1}\{T_{i}(1 - t) = t \})] \\
    &= \mathbb{E}[g(Y_{i}(t), z, X_{i}) \mathbbm{1}\{ T_{i}(t) = t \} ] - \mathbb{E}[g(Y_{i}(t), z, X_{i}) \mathbbm{1}\{ T_{i}(1 - t) = t \} ] \\
    &= \mathbb{E}[g(Y_{i}(t), z, X_{i}) \mathbbm{1}\{ T_{i}(t) = t \} \mid Z_{i} = t ] - \mathbb{E}[g(Y_{i}(t), z, X_{i}) \mathbbm{1}\{ T_{i}(1 - t) = t \} \mid Z_{i} = 1 - t ] \\
    &= \mathbb{E}[g(Y_{i}, z, X_{i}) \mathbbm{1}\{ T_{i} = t \} \mid Z_{i} = t ] - \mathbb{E}[g(Y_{i}, z, X_{i}) \mathbbm{1}\{ T_{i} = t \} \mid Z_{i} = 1 - t ],
\end{align*}
where the first and the second equality uses IV monotonicity in Assumption~\ref{assume_potential_model}, and the fourth equality uses IV exogeneity in Assumption~\ref{assume_potential_model}. \qed

\subsection{Proof of Proposition~\ref{thm_kappa_g_t_x_cond_y}}

For $\mathbb{E}[g(T_{i}, X_{i}) \mid Y_{i}(t) = y, T_{i}(1) > T_{i}(0)]$ where $t, y \in \{0, 1\}$:
\begin{align*}
    &\mathbb{E}[g(T_{i}, X_{i}) \mid Y_{i}(t) = y, T_{i}(1) > T_{i}(0)] \\
    &= \frac{\mathbb{E}[g(T_{i}, X_{i}) \mathbbm{1}\{ Y_{i}(t) = y \} \mathbbm{1}\{T_{i}(1) > T_{i}(0) \}]}{\mathbb{P}[Y_{i}(t) = y, T_{i}(1) > T_{i}(0)]} \\
    &= \frac{\sum_{z \in \{0, 1\}} \mathbb{E}[g(z, X_{i}) \mathbbm{1}\{ Y_{i}(t) = y \} \mathbbm{1}\{T_{i}(1) > T_{i}(0) \}] \mathbb{P}[Z_{i} = z]}{\mathbb{P}[Y_{i}(t) = y, T_{i}(1) > T_{i}(0)]},
\end{align*}
where the second equality uses the fact that $T_{i} = Z_{i}$ among compliers and IV independence assumption.

$\mathbb{E}[g(z, X_{i}) \mathbbm{1}\{ Y_{i}(t) = y \} \mathbbm{1}\{T_{i}(1) > T_{i}(0) \}]$ is identified by:
\begin{align*}
    &\mathbb{E}[g(z, X_{i}) \mathbbm{1}\{ Y_{i}(t) = y \} \mathbbm{1}\{T_{i}(1) > T_{i}(0) \}] \\
    &= \mathbb{E}[g(z, X_{i}) \mathbbm{1}\{Y_{i} =  y\} \mathbbm{1}\{ T_{i} = t \} \mid Z_{i} = t ] - \mathbb{E}[g(z, X_{i}) \mathbbm{1}\{Y_{i} =  y\} \mathbbm{1}\{ T_{i} = t \} \mid Z_{i} = 1 - t ],
\end{align*}
which follows from applying a result shown in Theorem~\ref{thm_kappa_g_y_t_x} by defining $g(Y_{i}(t), z, X_{i}) = g(z, X_{i}) \mathbbm{1}\{ Y_{i}(t) = y \}$.

$\mathbb{P}[Y_{i}(t) = y, T_{i}(1) > T_{i}(0)]$ is identified by:
\begin{align*}
    &\mathbb{P}[Y_{i}(t) = y, T_{i}(1) > T_{i}(0)] \\
    &= \mathbb{E}[\mathbbm{1}\{ Y_{i} = y \} \mathbbm{1}\{ T_{i} = t \} \mid Z_{i} = t] - \mathbb{E}[\mathbbm{1}\{ Y_{i} = y \} \mathbbm{1}\{ T_{i} = t \} \mid Z_{i} = 1- t],
\end{align*}
which follows from applying a result shown in Theorem~\ref{thm_kappa_g_y_t_x} by defining $g(Y_{i}(t), z, X_{i}) = \mathbbm{1}\{ Y_{i}(t) = y \}$. \qed

\subsection{Proof of Theorem~\ref{thm_comliance_persuasion_kappa_x_t}}

The results for $\mathbb{E}[g(T_{i}, X_{i}) \mid Y_{i}(1) = Y_{i}(0) = 1, T_{i}(1) > T_{i}(0)]$ and $\mathbb{E}[g(T_{i},X_{i}) \mid Y_{i}(1) = Y_{i}(0) = 0, T_{i}(1) > T_{i}(0)]$ follow immediately by observing that Assumption~\ref{assume_potential_model} implies that $[Y_{i}(1) = Y_{i}(0) = 1] = [Y_{i}(0) = 1]$ and $[Y_{i}(1) = Y_{i}(0) = 0] = [Y_{i}(1) = 0]$ and applying Proposition~\ref{thm_kappa_g_t_x_cond_y}.

For $\mathbb{E}[g(T_{i},X_{i}) \mid Y_{i}(1) = 1, Y_{i}(0) = 0, T_{i}(1) > T_{i}(0)]$:
\begin{align*}
    &\mathbb{E}[g(T_{i},X_{i}) \mid Y_{i}(1) = 1, Y_{i}(0) = 0, T_{i}(1) > T_{i}(0)] \\
    &= \sum_{z \in \{0, 1\}} \left( \frac{\mathbb{E}[g(z, X_{i}) \mathbbm{1}\{ Y_{i}(1) = 1, Y_{i}(0) = 0, T_{i}(1) > T_{i}(0) \}]}{\mathbb{P}[Y_{i}(1) = 1, Y_{i}(0) = 0, T_{i}(1) > T_{i}(0)]} \right) \mathbb{P}[Z_{i} = z],
\end{align*}
where the equality uses the fact that $T_{i} = Z_{i}$ among compliers and the IV independence assumption.

For $\mathbb{E}[g(z, X_{i}) \mathbbm{1}\{ Y_{i}(1) = 1, Y_{i}(0) = 0, T_{i}(1) > T_{i}(0) \}]$:
\begin{align*}
    &\mathbb{E}[g(z, X_{i}) \mathbbm{1}\{Y_{i}(1) = 1, Y_{i}(0) = 0, T_{i}(1) > T_{i}(0) \}] \\
    &= \mathbb{E}[g(z, X_{i}) (Y_{i}(1) - Y_{i}(0)) (T_{i}(1) - T_{i}(0)) ] \\
    &= \mathbb{E}[g(z, X_{i}) (T_{i}(1) Y_{i}(1) + (1 - T_{i}(1)) Y_{i}(0)) ] \\
    &\quad - \mathbb{E}[g(z, X_{i}) (T_{i}(0) Y_{i}(1) + (1 - T_{i}(0)) Y_{i}(0)) ] \\
    &= \mathbb{E}[g(z, X_{i}) (T_{i}(1) Y_{i}(1) + (1 - T_{i}(1)) Y_{i}(0))  \mid Z_{i} = 1 ] \\
    &\quad - \mathbb{E}[g(z, X_{i}) (T_{i}(0) Y_{i}(1) + (1 - T_{i}(0))  Y_{i}(0)) \mid Z_{i} = 0 ] \\
    &= \mathbb{E}[g(z, X_{i}) (T_{i} Y_{i}(1) + (1 - T_{i}) Y_{i}(0))  \mid Z_{i} = 1 ] \\
    &\quad - \mathbb{E}[g(z, X_{i}) (T_{i} Y_{i}(1) + (1 - T_{i})  Y_{i}(0)) \mid Z_{i} = 0 ] \\
    &= \mathbb{E}[g(z, X_{i}) \mathbbm{1}\{Y_{i} = 1\} \mid Z_{i} = 1 ] - \mathbb{E}[g(X_{i}) \mathbbm{1}\{Y_{i} = 1\} \mid Z_{i} = 0 ],
\end{align*}
where the third equality uses the IV exogeneity in Assumption~\ref{assume_potential_model}.

The identification of $\mathbb{P}[Y_{i}(1) = 1, Y_{i}(0) = 0, T_{i}(1) > T_{i}(0)]$ follows by using the identification results for $\mathbb{E}[g(z, X_{i}) \mathbbm{1}\{ Y_{i}(1) = 1, Y_{i}(0) = 0, T_{i}(1) > T_{i}(0) \}]$ and define $g(z, X_{i}) = 1$. \qed

\subsection{Proof of Proposition~\ref{thm_profile_persudaded_types}}

First, consider $\mathbb{E}[g(X_{i}, T_{i}) \mathbbm{1}\{ Y_{i}(1) = 1 \} \mid Y_{i}(0) = 1, T_{i}(1) > T_{i}(0) ]$ and $\mathbb{E}[g(X_{i}, T_{i}) \mathbbm{1}\{ Y_{i}(0) = 0 \} \mid Y_{i}(1) = 0, T_{i}(1) > T_{i}(0) ]$. For $t \in \{0, 1\}$:
\begin{align*}
    &\mathbb{E}[g(X_{i}, T_{i}) \mathbbm{1}\{ Y_{i}(t) = t \} \mid Y_{i}(1 - t) = t, T_{i}(1) > T_{i}(0) ] = \mathbb{E}[g(X_{i}, T_{i}) \mid Y_{i}(1 - t) = t, T_{i}(1) > T_{i}(0) ],
\end{align*}
where the equality follows from the outcome monotonicity assumption in Assumption~\ref{assume_potential_model}.

Second, consider $\mathbb{E}[ g(X_{i}, T_{i}) \mathbbm{1}\{ Y_{i}(1) = 1 \} \mid Y_{i}(0) = 0, T_{i}(1) > T_{i}(0) ]$ and $\mathbb{E}[ g(X_{i}, T_{i}) \mathbbm{1}\{ Y_{i}(0) = 0 \} \mid Y_{i}(1) = 1, T_{i}(1) > T_{i}(0) ]$. For $t \in \{0, 1\}$:
\begin{align*}
    &\mathbb{E}[ g(X_{i}, T_{i}) \mathbbm{1}\{ Y_{i}(1 - t) = 1 - t \} \mid Y_{i}(t) = t, T_{i}(1) > T_{i}(0) ] \\
    &= \frac{\mathbb{E}[ g(X_{i}, T_{i}) \mathbbm{1}\{ Y_{i}(1 - t) = 1 - t, Y_{i}(t) = t, T_{i}(1) > T_{i}(0) \}]}{\mathbb{P}[Y_{i}(t) = t, T_{i}(1) > T_{i}(0)]} \\
    &= \sum_{z \in \{0, 1\}} \left( \frac{\mathbb{E}[g(X_{i}, z) \mathbbm{1}\{ Y_{i} = 1 \} \mid Z_{i} = 1] - \mathbb{E}[g(X_{i}, z) \mathbbm{1}\{ Y_{i} = 1 \} \mid Z_{i} = 0]}{\mathbb{P}[Y_{i} = t, T_{i} = t \mid Z_{i} = t] - \mathbb{P}[Y_{i} = t, T_{i} = t \mid Z_{i} = 1 - t]} \right) \mathbb{P}[Z_{i} = z],
\end{align*}
where the second equality uses Lemma~\ref{corollary_marginal_potential} and a result in Theorem~\ref{thm_comliance_persuasion_kappa_x_t}.

Finally, consider $\mathbb{E}[g(X_{i}, T_{i}) \mathbbm{1}\{ Y_{i}(1) = 0 \} \mid Y_{i}(0) = 0, T_{i}(1) > T_{i}(0)]$ and $\mathbb{E}[g(X_{i}, T_{i}) \mathbbm{1}\{ Y_{i}(0) = 1 \} \mid Y_{i}(1) = 1, T_{i}(1) > T_{i}(0)]$. For $t \in \{0, 1\}$:
\begin{align*}
    &\mathbb{E}[g(X_{i}) \mathbbm{1}\{ Y_{i}(1 - t) = t \} \mid Y_{i}(t) = t, T_{i}(1) > T_{i}(0)] \\
    &= \frac{\mathbb{E}[g(X_{i}, T_{i}) \mathbbm{1}\{ Y_{i}(1 - t) = t, Y_{i}(t) = t, T_{i}(1) > T_{i}(0) \} ]}{\mathbb{P}[Y_{i}(t) = t, T_{i}(1) > T_{i}(0)]} \\
    &= \frac{\mathbb{E}[g(X_{i}, T_{i}) \mathbbm{1}\{ Y_{i}(1 - t) = t, T_{i}(1) > T_{i}(0) \} ]}{\mathbb{P}[Y_{i}(t) = t, T_{i}(1) > T_{i}(0)]} \\
    &= \sum_{z \in \{0, 1\}} \left( \frac{\mathbb{E}[ g(X_{i}, z) \mathbbm{1}\{ Y_{i} = t, T_{i} = 1 - t \} \mid Z_{i} = 1 - t ] - \mathbb{E}[ g(X_{i}, z) \mathbbm{1}\{ Y_{i} = t, T_{i} = 1 - t \} \mid Z_{i} = t ]}{\mathbb{P}[Y_{i} = t, T_{i} = t \mid Z_{i} = t] - \mathbb{P}[Y_{i} = t, T_{i} = t \mid Z_{i} = 1 - t]} \right) \mathbb{P}[Z_{i} = z],
\end{align*}
where the second eqaulity uses the monotone treatment response assumption in Assumption~\ref{assume_potential_model}, the third equality uses Lemma~\ref{corollary_marginal_potential} and a result in Theorem~\ref{thm_comliance_persuasion_kappa_x_t}. \qed

\subsection{Proof of Proposition~\ref{thm_equiv_theta_dk_local}}

Recall the formulas of the approximated $\tilde{\theta}_{\text{DK}}$ and the identified $\theta_{\text{local}}$ from Theorem 6 in \cite{jun2018identifying}:
\begin{align*}
    \tilde{\theta}_{\text{DK}} &= \frac{\mathbb{P}[Y_{i} = 1 \mid Z_{i} = 1] - \mathbb{P}[Y_{i} = 1 \mid Z_{i} = 0]}{(\mathbb{P}[T_{i} = 1 \mid Z_{i} = 1] - \mathbb{P}[T_{i} = 1 \mid Z_{i} = 0]) \times (1 - \mathbb{P}[Y_{i} = 1 \mid Z_{i} = 0])} \\
    \theta_{\text{local}} &= \frac{\mathbb{P}[Y_{i} = 1 \mid Z_{i} = 1] - \mathbb{P}[Y_{i} = 1 \mid Z_{i} = 0]}{\mathbb{P}[Y_{i} = 0, T_{i} = 0 \mid Z_{i} = 0] - \mathbb{P}[Y_{i} =0, T_{i} = 0 \mid Z_{i} = 1]},
\end{align*}
thus, $\tilde{\theta}_{\text{DK}} = \theta_{\text{local}}$ if and only if:
\begin{align}
    &(\mathbb{P}[T_{i} = 1 \mid Z_{i} = 1] - \mathbb{P}[T_{i} = 1 \mid Z_{i} = 0]) \times \mathbb{P}[Y_{i} = 0 \mid Z_{i} = 0] \nonumber \\
    &= \mathbb{P}[Y_{i} = 0, T_{i} = 0 \mid Z_{i} = 0] - \mathbb{P}[Y_{i} =0, T_{i} = 0 \mid Z_{i} = 1]. \label{thm_equiv_theta_denom}
\end{align}

Consider the first case in which there is non-compliance in the control group, i.e., $\mathbb{P}[T_{i} = 1 \mid Z_{i} = 1] = 1$. In this case, there is no never-taker. Then, for the denominator of $\tilde{\theta}_{\text{DK}}$:
\begin{align*}
    &(\mathbb{P}[T_{i} = 1 \mid Z_{i} = 1] - \mathbb{P}[T_{i} = 1 \mid Z_{i} = 0]) \times (1 - \mathbb{P}[Y_{i} = 1 \mid Z_{i} = 0]) \\
    &= (1 - \mathbb{P}[T_{i} = 1 \mid Z_{i} = 0]) \times (\mathbb{P}[Y_{i} = 0 \mid Z_{i} = 0]) \\
    &=\mathbb{P}[T_{i} = 0 \mid Z_{i} = 0] \times (\mathbb{P}[Y_{i} = 0, T_{i} = 0 \mid Z_{i} = 0] + \mathbb{P}[Y_{i} = 0, T_{i} = 1 \mid Z_{i} = 0]) \\
    &= \mathbb{P}[T_{i}(0) = 0] \times (\mathbb{P}[Y_{i}(0) = 0, T_{i}(0) = 0] + \mathbb{P}[Y_{i}(1) = 0, T_{i}(0) = 1]),
\end{align*}
where the first equality uses the assumption that there is non-compliance in the control group. For the denominator of $\tilde{\theta}_{\text{DK}}$, by the assumption that there is non-compliance in the control group:
\begin{align*}
    &\mathbb{P}[Y_{i} = 0, T_{i} = 0 \mid Z_{i} = 0] - \mathbb{P}[Y_{i} =0, T_{i} = 0 \mid Z_{i} = 1] \\
    &= \mathbb{P}[Y_{i} = 0, T_{i} = 0 \mid Z_{i} = 0] \\
    &= \mathbb{P}[Y_{i}(0) = 0, T_{i}(0) = 0].
\end{align*}
Thus, by Equation~\ref{thm_equiv_theta_denom}, $\tilde{\theta}_{\text{DK}} = \theta_{\text{local}}$ if and only if:
\begin{align*}
    &\mathbb{P}[Y_{i}(0) = 0, T_{i}(0) = 0] = \mathbb{P}[T_{i}(0) = 0] \times (\mathbb{P}[Y_{i}(0) = 0, T_{i}(0) = 0] + \mathbb{P}[Y_{i}(1) = 0, T_{i}(0) = 1]) \\
    &\Leftrightarrow \mathbb{P}[T_{i}(0) = 1] \times \mathbb{P}[Y_{i}(0) = 0, T_{i}(0) = 0] = \mathbb{P}[T_{i}(0) = 0] \times \mathbb{P}[Y_{i}(1) = 0, T_{i}(0) = 1] \\
    &\Leftrightarrow \mathbb{P}[Y_{i}(0) = 0 \mid T_{i}(0) = 0] = \mathbb{P}[Y_{i}(1) = 0 \mid T_{i}(0) = 1].
\end{align*}

Consider the second case in which there is non-compliance in the treatment group, i.e., $\mathbb{P}[T_{i} = 0 \mid Z_{i} = 0] = 1$. In this case, there is no always-taker. Then, for the denominator of $\tilde{\theta}_{\text{DK}}$:
\begin{align*}
    &(\mathbb{P}[T_{i} = 1 \mid Z_{i} = 1] - \mathbb{P}[T_{i} = 1 \mid Z_{i} = 0]) \times (1 - \mathbb{P}[Y_{i} = 1 \mid Z_{i} = 0]) \\ 
    &= \mathbb{P}[T_{i} = 1 \mid Z_{i} = 1] \times \mathbb{P}[Y_{i} = 0, T_{i} = 0 \mid Z_{i} = 0] \\
    &= \mathbb{P}[Y_{i} = 0, T_{i} = 0 \mid Z_{i} = 0] - \mathbb{P}[T_{i} = 0 \mid Z_{i} = 1] \times \mathbb{P}[Y_{i} = 0, T_{i} = 0 \mid Z_{i} = 0], 
\end{align*}
where the first equality uses the assumption that there is non-compliance in the treatment group. Thus, by Equation~\ref{thm_equiv_theta_denom}, $\tilde{\theta}_{\text{DK}} = \theta_{\text{local}}$ if and only if:
\begin{align*}
    &\mathbb{P}[Y_{i} = 0, T_{i} = 0 \mid Z_{i} = 0] - \mathbb{P}[Y_{i} =0, T_{i} = 0 \mid Z_{i} = 1] \\
    &= \mathbb{P}[Y_{i} = 0, T_{i} = 0 \mid Z_{i} = 0] - \mathbb{P}[T_{i} = 0 \mid Z_{i} = 1] \times \mathbb{P}[Y_{i} = 0, T_{i} = 0 \mid Z_{i} = 0] \\
    &\Leftrightarrow \mathbb{P}[Y_{i} =0, T_{i} = 0 \mid Z_{i} = 1] = \mathbb{P}[T_{i} = 0 \mid Z_{i} = 1] \times \mathbb{P}[Y_{i} = 0, T_{i} = 0 \mid Z_{i} = 0] \\
    &\Leftrightarrow \mathbb{P}[Y_{i}(0) =0, T_{i}(1) = 0] = \mathbb{P}[T_{i}(1) = 0] \times \mathbb{P}[Y_{i}(0) = 0, T_{i}(0) = 0] \\
    &\Leftrightarrow \mathbb{P}[Y_{i}(0) =0 \mid T_{i}(1) = 0] = \mathbb{P}[Y_{i}(0) = 0] \\
    &\Leftrightarrow Y_{i}(0) \independent T_{i}(1),
\end{align*}
where the third line uses the assumption that $\mathbb{P}[T_{i}(0) = 0] = 1$. \qed

\subsection{Proof of Proposition~\ref{claim_local_untreat}}

Note that among compliers, $T_{i} = Z_{i}$. Now the desired result follows immediately by observing that $Z_{i}$ is exogenous assumed in Assumption~\ref{assume_potential_model} and using Theorem 6 in \cite{jun2018identifying}. \qed

\subsection{Proof of Proposition~\ref{prop_sharp_test}}

We first show that if Assumptions 1, 2, 4, and 5 in Assumption~\ref{assume_potential_model} hold, then, there exists $\mathbf{p} \geq \mathbf{0}$ such that $A_{\text{obs}} \mathbf{p} = \mathbf{b}$ for all measurable set $A$. Consider a measurable set $A$ and $t, y \in \{0, 1\}^{2}$. To ensure that the joint distribution of the potential outcomes, potential treatments, and covariates is valid, we need the following:
\begin{align*}
    &\mathbb{P}[Y_{i}(0) = y_0, Y_{i}(1) = y_1, T_{i}(0) = t_0, T_{i}(1) = t_1, X_{i} \in A] \geq 0, \text{where $y_0, y_1, t_0, t_1 \in \{0, 1\}^{4}$} \\
    &\sum_{y_0, y_1, t_0, t_1 \in \{0, 1\}^{4}} \mathbb{P}[Y_{i}(0) = y_0, Y_{i}(1) = y_1, T_{i}(0) = t_0, T_{i}(1) = t_1, X_{i} \in A] = \mathbb{P}[X_{i} \in A].
\end{align*}
Then, by Assumptions 1 and 2 in Assumption~\ref{assume_potential_model}:
\begin{align*}
    &\mathbb{P}[Y_{i} = y, T_{i} = t, X_{i} \in A \mid Z_{i} = z] \\
    &= \mathbb{P}[Y_{i}(t) = y, T_{i}(z) = t, X_{i} \in A] \\
    &= \mathbb{P}[Y_{i}(t) = y, Y_{i}(1 - t) = y, T_{i}(z) = t, T_{i}(1 - z) = t, X_{i} \in A] \\
    &\quad + \mathbb{P}[Y_{i}(t) = y, Y_{i}(1 - t) = y, T_{i}(z) = t, T_{i}(1 - z) = 1 - t, X_{i} \in A] \\
    &\quad + \mathbb{P}[Y_{i}(t) = y, Y_{i}(1 - t) = 1 - y, T_{i}(z) = t, T_{i}(1 - z) = t, X_{i} \in A] \\
    &\quad + \mathbb{P}[Y_{i}(t) = y, Y_{i}(1 - t) = 1 - y, T_{i}(z) = t, T_{i}(1 - z) = 1 - t, X_{i} \in A].
\end{align*}
Furthermore, Assumptions 4 and 5 in Assumption~\ref{assume_potential_model} implies the following:
\begin{align*}
    &\mathbb{P}[Y_{i}(0) = 1, Y_{i}(1) = 0, T_{i}(z) = t, T_{i}(1 - z) = 1 - t, X_{i} \in A] = 0, \\
    &\mathbb{P}[Y_{i}(t) = y, Y_{i}(1 - t) = 1 - y, T_{i}(0) = 1, T_{i}(1) = 0, X_{i} \in A] = 0,
\end{align*}
where $t, y, z \in \{0, 1\}^{3}$. Combining the above restrictions yields a system of linear equations $A_{\text{obs}} \mathbf{p} = \mathbf{b}$ such that $\mathbf{p} \geq \mathbf{0}$.

For the other direction, consider the following joint distribution of potential outcomes, potential treatments, covariates, and the instrument, denoted by $\text{Pr}$:
\begin{align*}
&\text{Pr}[\tilde{Y}_{i}(0) = y_0, \tilde{Y}_{i}(1) = y_1, \tilde{T}_{i}(0) = t_0, \tilde{T}_{i}(1) = t_1, X_{i} \in A, Z_{i} = z] \\
&= \mathbb{P}[\tilde{Y}_{i}(0) = y_0, \tilde{Y}_{i}(1) = y_1, \tilde{T}_{i}(0) = t_0, \tilde{T}_{i}(1) = t_1, X_{i} \in A] \times \mathbb{P}[Z_{i} = z],
\end{align*}
where $y_0, y_1, t_0, t_1, z \in \{0, 1\}^{5}$, and $A$ is a measurable set. $\mathbb{P}[Z_{i} = z]$ is the observed distribution of $Z_{i}$. $\mathbb{P}[\tilde{Y}_{i}(0) = y_0, \tilde{Y}_{i}(1) = y_1, \tilde{T}_{i}(0) = t_0, \tilde{T}_{i}(1) = t_1, X_{i} \in A]$ satisfies the linear constraints, $A_{\text{obs}} \mathbf{p} = \mathbf{b}$ such that $\mathbf{p} \geq \boldsymbol{0}$. The linear constraints ensure that $\mathbb{P}[\tilde{Y}_{i}(0) = y_0, \tilde{Y}_{i}(1) = y_1, \tilde{T}_{i}(0) = t_0, \tilde{T}_{i}(1) = t_1, X_{i} \in A]$ is a valid probability distribution (additive, non-negative, and add up to one). Thus, by the product measure theorem, $\text{Pr}[\tilde{Y}_{i}(0) = y_0, \tilde{Y}_{i}(1) = y_1, \tilde{T}_{i}(0) = t_0, \tilde{T}_{i}(1) = t_1, X_{i} \in A, Z_{i} = z]$ is a valid distribution.

With this construction, the IV independence assumption is satisfied automatically. Furthermore, since the distribution of $(\tilde{Y}_{i}(0) = y_0, \tilde{Y}_{i}(1) = y_1, \tilde{T}_{i}(0) = t_0, \tilde{T}_{i}(1) = t_1, X_{i} \in A)$ satisfies $A_{\text{obs}} \mathbf{p} = \mathbf{b}$ such that $\mathbf{p} \geq \mathbf{0}$, the IV monotonicity assumption and the monotone treatment response assumption are satisfied.

Finally, we check that the constructed distribution induces observables matching the data. For $t, y, z \in \{0, 1\}^{3}$, and a measurable set $A$:
\begin{align*}
    &\mathbb{P}[Y_{i} = y, T_{i} = t \mid Z_{i} = z] \\
    &= \mathbb{P}[\tilde{Y_{i}}(t) = y, \tilde{Y_{i}}(1 - t) = y, \tilde{T_{i}}(z) = t, \tilde{T_{i}}(1 - z) = t, X_{i} \in A] \\
    &\quad + \mathbb{P}[\tilde{Y_{i}}(t) = y, \tilde{Y_{i}}(1 - t) = y, \tilde{T_{i}}(z) = t, \tilde{T_{i}}(1 - z) = 1 - t, X_{i} \in A] \\
    &\quad + \mathbb{P}[\tilde{Y_{i}}(t) = y, \tilde{Y_{i}}(1 - t) = 1 - y, \tilde{T_{i}}(z) = t, \tilde{T_{i}}(1 - z) = t, X_{i} \in A] \\
    &\quad + \mathbb{P}[\tilde{Y_{i}}(t) = y, \tilde{Y_{i}}(1 - t) = 1 - y, \tilde{T_{i}}(z) = t, \tilde{T_{i}}(1 - z) = 1 - t, X_{i} \in A] \\
    &= \text{Pr}[\tilde{Y_{i}}(t) = y, \tilde{Y_{i}}(1 - t) = y, \tilde{T_{i}}(z) = t, \tilde{T_{i}}(1 - z) = t, X_{i} \in A] \\
    &\quad + \text{Pr}[\tilde{Y_{i}}(t) = y, \tilde{Y_{i}}(1 - t) = y, \tilde{T_{i}}(z) = t, \tilde{T_{i}}(1 - z) = 1 - t, X_{i} \in A] \\
    &\quad + \text{Pr}[\tilde{Y_{i}}(t) = y, \tilde{Y_{i}}(1 - t) = 1 - y, \tilde{T_{i}}(z) = t, \tilde{T_{i}}(1 - z) = t, X_{i} \in A] \\
    &\quad + \text{Pr}[\tilde{Y_{i}}(t) = y, \tilde{Y_{i}}(1 - t) = 1 - y, \tilde{T_{i}}(z) = t, \tilde{T_{i}}(1 - z) = 1 - t, X_{i} \in A] \\
    &= \text{Pr}[\tilde{Y_{i}}(t) = y, \tilde{T_{i}}(z) = t, X_{i} \in A \mid Z_{i} = z] \\
    &= \text{Pr}[\tilde{Y_{i}} = y, \tilde{T_{i}} = t, X_{i} \in A \mid Z_{i} = z],
\end{align*}
where the first equality uses the fact $\mathbb{P}[\tilde{Y}_{i}(0) = y_0, \tilde{Y}_{i}(1) = y_1, \tilde{T}_{i}(0) = t_0, \tilde{T}_{i}(1) = t_1, X_{i} \in A]$ satisfies the linear constraints, $A_{\text{obs}} \mathbf{p} = \mathbf{b}$ such that $\mathbf{p} \geq \boldsymbol{0}$, the second equality follows from:
\begin{align*}
&\text{Pr}[\tilde{Y}_{i}(0) = y_0, \tilde{Y}_{i}(1) = y_1, \tilde{T}_{i}(0) = t_0, \tilde{T}_{i}(1) = t_1, X_{i} \in A] \\
&= \sum_{z \in \{0, 1\}} \text{Pr}[\tilde{Y}_{i}(0) = y_0, \tilde{Y}_{i}(1) = y_1, \tilde{T}_{i}(0) = t_0, \tilde{T}_{i}(1) = t_1, X_{i} \in A, Z_{i} = z] \\
&= \sum_{z \in \{0, 1\}} \mathbb{P}[\tilde{Y}_{i}(0) = y_0, \tilde{Y}_{i}(1) = y_1, \tilde{T}_{i}(0) = t_0, \tilde{T}_{i}(1) = t_1, X_{i} \in A] \times \mathbb{P}[Z_{i} = z] \\
&= \mathbb{P}[\tilde{Y}_{i}(0) = y_0, \tilde{Y}_{i}(1) = y_1, \tilde{T}_{i}(0) = t_0, \tilde{T}_{i}(1) = t_1, X_{i} \in A],
\end{align*}
the third and fourth equality uses the fact that the constructed distribution for $(\tilde{Y}_{i}(0) = y_0, \tilde{Y}_{i}(1) = y_1, \tilde{T}_{i}(0) = t_0, \tilde{T}_{i}(1) = t_1, X_{i} \in A)$ satisfies IV independence, IV monotonicity, and monotone treatment response. \qed

\subsection{Proof of Proposition~\ref{prop_unidentifiable_joint_po_nonbinary}}

Note that the marginal distribution of potential outcomes among compliers is point identified (\citealpAppx{imbens1997estimating}, \citealpAppx{abadie2003semiparametric}). Moreover, we can rewrite the marginal distribution of potential outcomes among compliers as a system of linear equations of the joint distribution of potential outcomes among compliers:
\begin{align*}
\begin{bmatrix}
1 & 1 & 1 & 0 & 0 & 0 \\
0 & 0 & 0 & 1 & 1 & 0 \\
0 & 0 & 0 & 0 & 0 & 1 \\
1 & 0 & 0 & 0 & 0 & 0 \\
0 & 1 & 0 & 1 & 0 & 0 \\
0 & 0 & 1 & 0 & 1 & 1 \\
1 & 1 & 1 & 1 & 1 & 1
\end{bmatrix}
\begin{bmatrix}
\mathbb{P}[Y_{i}(0) = -1, Y_{i}(1) = -1 | T_{i}(1) > T_{i}(0)] \\
\mathbb{P}[Y_{i}(0) = -1, Y_{i}(1) = 0 | T_{i}(1) > T_{i}(0)] \\
\mathbb{P}[Y_{i}(0) = -1, Y_{i}(1) = 1 | T_{i}(1) > T_{i}(0)] \\
\mathbb{P}[Y_{i}(0) = 0, Y_{i}(1) = 0 | T_{i}(1) > T_{i}(0)] \\
\mathbb{P}[Y_{i}(0) = 0, Y_{i}(1) = 1 | T_{i}(1) > T_{i}(0)] \\
\mathbb{P}[Y_{i}(0) = 1, Y_{i}(1) = 1 | T_{i}(1) > T_{i}(0)]
\end{bmatrix} = 
\begin{bmatrix}
\mathbb{P}[Y_{i}(0) = -1 | T_{i}(1) > T_{i}(0)] \\
\mathbb{P}[Y_{i}(0) = 0 | T_{i}(1) > T_{i}(0)] \\
\mathbb{P}[Y_{i}(0) = 1 | T_{i}(1) > T_{i}(0)] \\
\mathbb{P}[Y_{i}(1) = -1 | T_{i}(1) > T_{i}(0)] \\
\mathbb{P}[Y_{i}(1) = 0 | T_{i}(1) > T_{i}(0)] \\
\mathbb{P}[Y_{i}(1) = 1 | T_{i}(1) > T_{i}(0)] \\ 
1
\end{bmatrix},
\end{align*}
where the rank of the coefficient matrix is five. Thus, there is no unique solution to the system of linear equations above. \qed

\subsection{Proof of Corollary~\ref{corollary_local_persuasion_joint_multivalue_iv}}

The desired results follow immediately using the identical arguments in Lemma~\ref{corollary_marginal_potential} and Lemma~\ref{corollary_local_persuasion_joint}. \qed

\subsection{Proof of Corollary~\ref{corollary_joint_dist_po_conti_z}}

The desired result follows immediately by using the result in \citeAppx{heckman2005structural} and \citeAppx{carneiro2009estimating} and the monotone treatment response assumption in Assumption~\ref{assume_potential_model_multivalue_iv}. Since the argument is brief, we include it here for completeness.

Note that by the monotone treatment response assumption in Assumption~\ref{assumption_binary_t_y_conti_z} and the fact that $Y_{i}$ is binary:
\begin{align*}
    &\mathbb{P}[Y_{i}(1) = 1, Y_{i}(0) = 0 \mid V_{i} = v] = \mathbb{E}[Y_{i}(1) - Y_{i}(0) \mid V_{i} = v] \\
    &\mathbb{P}[Y_{i}(1) = Y_{i}(0) = 1 \mid V_{i} = v] = \mathbb{P}[Y_{i}(0) = 1 \mid V_{i} = v] \\
    &\mathbb{P}[Y_{i}(1) = Y_{i}(0) = 0 \mid V_{i} = v] = \mathbb{P}[Y_{i}(1) = 0 \mid V_{i} = v].
\end{align*}

To identify $\mathbb{E}[Y_{i}(1) - Y_{i}(0) \mid V_{i} = v]$, consider $\mathbb{E}[Y_{i} \mid V_{i} = v]$:
\begin{align*}
    \mathbb{E}[Y_{i} \mid V_{i} = v] &= \mathbb{E}[Y_{i}(0) \mid P(Z_{i}) = v] + \mathbb{E}[T_{i} (Y_{i}(1) - Y_{i}(0)) \mid P(Z_{i}) = v] \\
    &= \mathbb{E}[Y_{i}(0) \mid P(Z_{i}) = v] + \mathbb{E}[Y_{i}(1) - Y_{i}(0) \mid T_{i} = 1, P(Z_{i}) = v] \mathbb{P}[T_{i} = 1 \mid P(Z_{i}) = v] \\
    &= \mathbb{E}[Y_{i}(0) \mid P(Z_{i}) = v] + \mathbb{E}[Y_{i}(1) - Y_{i}(0) \mid V_{i} \leq v, P(Z_{i}) = v] \mathbb{P}[V_{i} \leq v \mid P(Z_{i}) = v] \\
    &= \mathbb{E}[Y_{i}(0)] + \mathbb{E}[Y_{i}(1) - Y_{i}(0) \mid V_{i} \leq v] v \\
    &= \mathbb{E}[Y_{i}(0)] + \mathbb{E}[(Y_{i}(1) - Y_{i}(0)) \mathbbm{1}\{ V_{i} \leq v \}] \\
    &= \mathbb{E}[Y_{i}(0)] + \mathbb{E}[\mathbbm{1}\{ V_{i} \leq v \} \mathbb{E}[Y_{i}(1) - Y_{i}(0) \mid V_{i} = u]] \\
    &= \mathbb{E}[Y_{i}(0)] + \int_{0}^{v} \mathbb{E}[Y_{i}(1) - Y_{i}(0) \mid V_{i} = u] du,
\end{align*}
where the third equality uses the selection equation in Assumption~\ref{assumption_binary_t_y_conti_z}, the fourth equality uses the independence of $Z_{i}$ and $V_{i} \sim U[0, 1]$ in Assumption~\ref{assumption_binary_t_y_conti_z}. Now the desired result follows immediately by differentiating both sides of the equation with respect to $v$.

To identify $\mathbb{P}[Y_{i}(0) = 1 \mid V_{i} = v]$, consider $(1 - v) \mathbb{E}[g(Y_{i}) \mid P(Z_{i}) = v, T_{i} = 0]$, where $g$ is a measurable map:
\begin{align*}
    (1 - v) \mathbb{E}[g(Y_{i}) \mid P(Z_{i}) = v, T_{i} = 0] &= (1 - v) \mathbb{E}[g(Y_{i}(0)) \mid V_{i} > v] \\
    &= \mathbb{E}[g(Y_{i}(0)) \mathbbm{1}\{ V_{i} > v \}] \\
    &= \mathbb{E}[\mathbbm{1}\{ V_{i} > v \} \mathbb{E}[g(Y_{i}(0)) \mid V_{i} = u]] \\
    &= \int_{v}^{1} \mathbb{E}[g(Y_{i}(0)) \mid V_{i} = u] du,
\end{align*}
where the first equality uses the selection equation in Assumption~\ref{assumption_binary_t_y_conti_z}, the fourth equality uses $V_{i} \sim U[0, 1]$ in Assumption~\ref{assumption_binary_t_y_conti_z}. Now the desired result follows immediately by differentiating both sides of the equation with respect to $v$ and defining $g$ as: $g(Y_{i}) = \mathbbm{1}\{ Y_{i} = 1 \}$.

To identify $\mathbb{P}[Y_{i}(1) = 0 \mid V_{i} = v]$, consider $v \mathbb{E}[g(Y_{i}) \mid P(Z_{i}) = v, T_{i} = 1]$, where $g$ is a measurable map:
\begin{align*}
    v \mathbb{E}[g(Y_{i}) \mid P(Z_{i}) = v, T_{i} = 1] &= v \mathbb{E}[g(Y_{i}(1)) \mid V_{i} \leq v] \\
    &= \mathbb{E}[g(Y_{i}(1)) \mathbbm{1}\{ V_{i} \leq v \}] \\
    &= \mathbb{E}[\mathbbm{1}\{ V_{i} \leq v \} \mathbb{E}[g(Y_{i}(1)) \mid V_{i} = u]] \\
    &= \int_{0}^{v} \mathbb{E}[g(Y_{i}(1)) \mid V_{i} = u] du,
\end{align*}
where the first equality uses the selection equation in Assumption~\ref{assumption_binary_t_y_conti_z}, the fourth equality uses $V_{i} \sim U[0, 1]$ in Assumption~\ref{assumption_binary_t_y_conti_z}. Now the desired result follows immediately by differentiating both sides of the equation with respect to $v$ and defining $g$ as: $g(Y_{i}) = \mathbbm{1}\{ Y_{i} = 0 \}$. \qed

\subsection{Proof of Corollary~\ref{corollary_profile_persuasion_binary_ty_discretez}}

The desired results follow immediately using the identical arguments in Theorem~\ref{thm_profile_persudaded_types}.

\subsection{Proof of Corollary~\ref{corollary_profile_persuasion_binary_ty_contiz}}

For $\mathbb{E}[g(X_{i}) \mid Y_{i}(1) = 1, Y_{i}(0) = 0, V_{i} = v]$:
\begin{align*}
    \mathbb{E}[g(X_{i}) \mid Y_{i}(1) = 1, Y_{i}(0) = 0, V_{i} = v] &= \frac{\mathbb{E}[g(X_{i}) \mathbbm{1}\{ Y_{i}(1) = 1, Y_{i}(0) = 0 \} \mid V_{i} = v]}{\mathbb{P}[Y_{i}(1) = 1, Y_{i}(0) = 0 \mid V_{i} = v]} \\
    &= \frac{\mathbb{E}[g(X_{i}) (Y_{i}(1) - Y_{i}(0)) \mid V_{i} = v]}{\mathbb{E}[Y_{i}(1) - Y_{i}(0) \mid V_{i} = v]} \\
    &= \frac{\frac{\partial}{\partial v} \mathbb{E}[g(X_{i}) Y_{i} \mid P(Z_{i}) = v]}{\frac{\partial}{\partial v} \mathbb{E}[Y_{i} \mid P(Z_{i}) = v]},
\end{align*}
where the second equality uses the monotone treatment response assumption, and the third equality uses the independence assumption in Assumption~\ref{assumption_binary_t_y_conti_z} and Corollary~\ref{corollary_joint_dist_po_conti_z}.

Now. consider $\mathbb{E}[g(X_{i}) \mid Y_{i}(1) = Y_{i}(0) = 1, V_{i} = v]$ and $\mathbb{E}[g(X_{i}) \mid Y_{i}(1) = Y_{i}(0) = 0, V_{i} = v]$. For $t \in \{0, 1 \}$:
\begin{align*}
    \mathbb{E}[g(X_{i}) \mid Y_{i}(1) = Y_{i}(0) = 1 - t, V_{i} = v] &= \mathbb{E}[g(X_{i}) \mid Y_{i}(t) = 1 - t, V_{i} = v] \\
    &= \frac{\mathbb{E}[g(X_{i}) \mathbbm{1}\{Y_{i}(t) = 1 - t\} \mid V_{i} = v]}{\mathbb{P}[Y_{i}(t) = 1 - t \mid V_{i} = v] },
\end{align*}
where the second equality uses the monotone treatment response assumption. Now the desired result follows immediately from the independence assumption in Assumption~\ref{assumption_binary_t_y_conti_z} and Corollary~\ref{corollary_joint_dist_po_conti_z}. \qed

\subsection{Proof of Proposition~\ref{prop_local_persuasion_abadie_atnt}}

For $\mathbb{E}[g(X_{i}) | Y_{i}(t) = y, T_{i}(1) = T_{i}(0) = t]$, where $t \in \{0, 1\}$ and $y \in \{0, 1\}$, we have the following:
\begin{align*}
    \mathbb{E}[g(X_{i}) | Y_{i}(t) = y, T_{i}(1) = T_{i}(0) = t] &= \mathbb{E}[g(X_{i}) | Y_{i}(t) = y, T_{i}(1 - t) = t] \\
    &= \mathbb{E}[g(X_{i}) | Y_{i}(t) = y, T_{i}(1 - t) = t, Z_{i} = 1 - t] \\   &= \mathbb{E}[g(X_{i}) | Y_{i} = y, T_{i} = t, Z_{i} = 1 - t],
\end{align*}
where the first equality uses the IV monotonicity assumption in Assumption~\ref{assume_potential_model}, the second equality uses the IV independence assumption in Assumption~\ref{assume_potential_model}. \qed

\subsection{A Glivenko-Cantelli Theorem for Conditional Cumulative Distribution Function}\label{appx_gelivenko_cantelli}

In fact, we can strengthen the statement in Appendix~\ref{appx_estimation_inference} from convergence in probability to almost sure convergence:
\begin{align*}
    \sup_{x \in \mathbb{R}} \left| \hat{\mathbb{P}}[ X_{i} \leq x \mid Y_{i}(0) = 0, T_{i}(1) > T_{i}(0) ] - \mathbb{P}[ X_{i} \leq x \mid Y_{i}(0) = 0, T_{i}(1) > T_{i}(0) ] \right| \xrightarrow[]{\text{a.s.}} 0.
\end{align*}
Moreover, the uniform convergence result follows immediately from the uniform convergence of the empirical conditional cumulative distribution function. Thus, we only provide a proof for the uniform convergence of the empirical conditional cumulative distribution function in this section.

\begin{theorem}
\rm Consider a pair of random variable $(X_{i}, Z_{i}): (\Omega, \mathcal{F}) \xrightarrow[]{} (\mathbb{R}^{2}, \mathcal{R}^{2})$, where $\mathcal{F}$ is a sigma field on the outcome space $\Omega$, and $\mathcal{R}^{2}$ denotes the Borel $\sigma$-algebra on $\mathbb{R}^{2}$. Let $A$ be a measurable set with $\mathbb{P}[Z_{i} \in A] \neq 0$. Then:
\begin{align*}
    \sup_{x \in \mathbb{R}} \left| \hat{\mathbb{P}}[X_{i} \leq x \mid Z_{i} \in A] - \mathbb{P}[X_{i} \leq x \mid Z_{i} \in A] \right| \xrightarrow[]{\text{a.s.}} 0,
\end{align*}
where $\hat{\mathbb{P}}[X_{i} \leq x \mid Z_{i} \in A] = \frac{\mathbb{E}_{n}[\mathbbm{1}\{ X_{i} \leq x, Z_{i} \in A \}]}{\mathbb{E}_{n}[\mathbbm{1}\{ Z_{i} \in A \}]}$ with $\mathbb{E}_{n}$ denotes sample average.
\end{theorem}

\begin{proof}
We first show that $\sup_{x \in \mathbb{R}} \left| \mathbb{E}_{n}[\mathbbm{1}\{ X_{i} \leq x, Z_{i} \in A \} ] - \mathbb{P}[X_{i} \leq x, Z_{i} \in A] \right| \xrightarrow[]{\text{a.s.}} 0$. For $1 \leq j \leq k - 1$, let $x_{j, k} = \inf\{y: \mathbb{P}[X_{i} \leq y, Z_{i} \in A] \geq \frac{j}{k} \mathbb{P}[Z_{i} \in A] \}$. Thus, by the Strong Law of Large Numbers, there exists $N_{k}$ such that if $n \geq N_{k}$, then:
\begin{align*}
    &\left|\mathbb{E}_{n}[\mathbbm{1}\{ Z_{i} \in A \}] - \mathbb{P}[Z_{i} \in A] \right| < \frac{\mathbb{P}[Z_{i} \in A]}{k}, \\
    &\left|\mathbb{E}_{n}[\mathbbm{1}\{ X_{i} \leq x_{j, k}, Z_{i} \in A \}] - \mathbb{P}[X_{i} \leq x_{j, k}, Z_{i} \in A] \right| < \frac{\mathbb{P}[Z_{i} \in A]}{k}, \\
    &\left|\mathbb{E}_{n}[\mathbbm{1}\{ X_{i} < x_{j, k}, Z_{i} \in A \}] - \mathbb{P}[X_{i} < x_{j, k}, Z_{i} \in A] \right| < \frac{\mathbb{P}[Z_{i} \in A]}{k},
\end{align*}
for $1 \leq j \leq k - 1$. With $x_{0, k} = - \infty$ and $x_{k, k} = \infty$, then the last two inequalities hold for $j = 0$ and $j = k$.

For $x \in (x_{j - 1, k}, x_{j, k})$ with $1 \leq j \leq k$ and $n \geq N_{k}$:
\begin{align*}
    \mathbb{E}_{n}[\mathbbm{1}\{X_{i} \leq x, Z_{i} \in A\}] &\leq \mathbb{E}_{n}[\mathbbm{1}\{X_{i} < x_{j, k}, Z_{i} \in A\}] \\
    &\leq \mathbb{E}[\mathbbm{1}\{ X_{i} < x_{j, k}, Z_{i} \in A \}] + \frac{\mathbb{P}[Z_{i} \in A]}{k} \\
    &\leq \mathbb{E}[\mathbbm{1}\{X_{i} \leq x_{j - 1, k}, Z_{i} \in A\}] + \frac{2 \mathbb{P}[Z_{i} \in A]}{k} \\
    &\leq \mathbb{E}[\mathbbm{1}\{ X_{i} \leq x, Z_{i} \in A \}] + \frac{2 \mathbb{P}[Z_{i} \in A\}]}{k}, \\
    \mathbb{E}_{n}[\mathbbm{1}\{X_{i} \leq x, Z_{i} \in A\}] &\geq \mathbb{E}_{n}[\mathbbm{1}\{X_{i} \leq x_{j - 1, k}, Z_{i} \in A\}] \\
    &\geq \mathbb{E}[\mathbbm{1}\{X_{i} \leq x_{j - 1, k}, Z_{i} \in A\}] - \frac{\mathbb{P}[Z_{i} \in A]}{k} \\
    &\geq \mathbb{E}[\mathbbm{1}\{X_{i} < x_{j, k}, Z_{i} \in A\}] - \frac{2 \mathbb{P}[Z_{i} \in A]}{k} \\
    &\geq \mathbb{E}[\mathbbm{1}\{X_{i} \leq x, Z_{i} \in A\}] - \frac{2 \mathbb{P}[Z_{i} \in A]}{k}.
\end{align*}
Thus, we conclude that $\sup_{x \in \mathbb{R}} \left| \mathbb{E}_{n}[\mathbbm{1}\{X_{i} \leq x, Z_{i} \in A\}] - \mathbb{P}[X_{i} \leq x, Z_{i} \in A] \right| \xrightarrow[]{\text{a.s.}} 0$.

For $\sup_{x \in \mathbb{R}} \left| \hat{\mathbb{P}}[X_{i} \leq x \mid Z_{i} \in A] - \mathbb{P}[X_{i} \leq x \mid Z_{i} \in A] \right|$:
\begin{align*}
\begin{split}
    &\sup_{x \in \mathbb{R}} \left| \hat{\mathbb{P}}[X_{i} \leq x \mid Z_{i} \in A] - \mathbb{P}[X_{i} \leq x \mid Z_{i} \in A] \right| \\
    &= \sup_{x \in \mathbb{R}} \left| \frac{\mathbb{E}_{n}[\mathbbm{1}\{ X_{i} \leq x, Z_{i} \in A \}]}{\mathbb{E}_{n}[\mathbbm{1}\{ Z_{i} \in A \}]} - \mathbb{P}[X_{i} \leq x \mid Z_{i} \in A] \right| \\
    &= \sup_{x \in \mathbb{R}} \left| \frac{\mathbb{E}_{n}[\mathbbm{1}\{ X_{i} \leq x, Z_{i} \in A \}]}{\mathbb{E}_{n}[\mathbbm{1}\{ Z_{i} \in A \}]} - \frac{\mathbb{E}_{n}[\mathbbm{1}\{ X_{i} \leq x, Z_{i} \in A \}]}{\mathbb{P}[\{ Z_{i} \in A \}]} + \frac{\mathbb{E}_{n}[\mathbbm{1}\{ X_{i} \leq x, Z_{i} \in A \}]}{\mathbb{P}[\{ Z_{i} \in A \}]} - \mathbb{P}[X_{i} \leq x \mid Z_{i} \in A] \right| \\
    &\leq \sup_{x \in \mathbb{R}} \left| \frac{\mathbb{E}_{n}[\mathbbm{1}\{ X_{i} \leq x, Z_{i} \in A \}]}{\mathbb{E}_{n}[\mathbbm{1}\{ Z_{i} \in A \}]} - \frac{\mathbb{E}_{n}[\mathbbm{1}\{ X_{i} \leq x, Z_{i} \in A \}]}{\mathbb{P}[\{ Z_{i} \in A \}]} \right| \\
    &\quad + \sup_{x \in \mathbb{R}} \left| \frac{\mathbb{E}_{n}[\mathbbm{1}\{ X_{i} \leq x, Z_{i} \in A \}]}{\mathbb{P}[\{ Z_{i} \in A \}]} - \mathbb{P}[X_{i} \leq x \mid Z_{i} \in A] \right|  \\
    &= \left| \frac{1}{\mathbb{E}_{n}[\mathbbm{1}\{ Z_{i} \in A \}]} - \frac{1}{\mathbb{P}[\{ Z_{i} \in A \}]} \right| \sup_{x \in \mathbb{R}} \left| \mathbb{E}_{n}[\mathbbm{1}\{ X_{i} \leq x, Z_{i} \in A \}] \right| \\
    &\quad + \frac{1}{\mathbb{P}[Z_{i} \in A]} \sup_{x \in \mathbb{R}} \left| \mathbb{E}_{n}[\mathbbm{1}\{ X_{i} \leq x, Z_{i} \in A \}] - \mathbb{P}[X_{i} \leq x, Z_{i} \in A] \right| \\
    &\leq \left| \frac{1}{\mathbb{E}_{n}[\mathbbm{1}\{ Z_{i} \in A \}]} - \frac{1}{\mathbb{P}[\{ Z_{i} \in A \}]} \right| + \frac{1}{\mathbb{P}[Z_{i} \in A]} \sup_{x \in \mathbb{R}} \left| \mathbb{E}_{n}[\mathbbm{1}\{ X_{i} \leq x, Z_{i} \in A \}] - \mathbb{P}[X_{i} \leq x, Z_{i} \in A] \right| \\
    &\xrightarrow[]{\text{a.s.}} 0,
\end{split}
\end{align*}
where the first inequality uses the triangle inequality, the second inequality uses the fact that:
\begin{align*}
    \sup_{x \in \mathbb{R}} \left| \mathbb{E}_{n}[\mathbbm{1}\{ X_{i} \leq x, Z_{i} \in A \}] \right| \leq 1,
\end{align*}
which holds by construction, and the last line uses the Strong Law of Large Numbers and the continuous mapping theorem. \qed
\end{proof}

\section{Identifiability of the Joint Distribution of Non-Binary Instruments or Outcomes}\label{appx_identify_nonbinary_z_y}

This section covers two potential directions for extending Lemma~\ref{corollary_local_persuasion_joint}. The first direction explores the positive outcomes that arise from utilizing a non-binary instrument to extend Lemma~\ref{corollary_local_persuasion_joint}. Following this, we delve into the negative outcomes associated with using a non-binary outcome to extend~\ref{corollary_local_persuasion_joint}.

\subsection{Non-Binary Instrument}\label{appx_identify_nonbinary_z}

Assumption~\ref{assume_potential_model} is adjusted to accommodate a discrete-valued instrument in two ways. Firstly, the IV monotonicity condition is crucially modified. With a discrete-valued instrument, the IV ``monotonicity" condition must be satisfied for each pair of instruments. That is, changing the instrument from $z$ to $z'$ will either encourage or discourage every individual from taking up the treatment. Secondly, the IV relevance assumption is also revised. In this case, at least one instrument value must lead to changes in selection behaviour. The formal statement of the revised assumption is now presented as Assumption~\ref{assume_potential_model_multivalue_iv}.

\begin{assumption}(Potential Outcome and Treatment Model with Discrete Valued Instrument)\label{assume_potential_model_multivalue_iv}
\begin{enumerate}
    \item Monotone treatment response: $Y_{i}(1) \geq Y_{i}(0)$ holds almost surely with $Y_{i}(0)$ and $Y_{i}(1)$ binary,
    \item Exclusion restriction: $Y_{i}(t, z) = Y_{i}(t)$, for $t, z \in \supp(T_{i}, Z_{i})$,
    \item Exogenous instrument: $Z_{i} \independent (Y_{i}(0), Y_{i}(1), T_{i}(0), T_{i}(1), X_{i})$,
    \item First stage: $\mathbb{P}[T_{i} = 1 | Z_{i} = z]$ is a non-trivial function of $z$,
    \item IV Monotonicity: either $T_{i}(z) \geq T_{i}(z')$ or $T_{i}(z) \leq T_{i}(z')$ holds almost surely for $z \neq z'$ with $z, z' \in \supp(Z_{i})$.
\end{enumerate}
\end{assumption}

With Assumption~\ref{assume_potential_model_multivalue_iv}, we can point identify the joint distribution of potential outcomes among each complier group. The intuition of the result is that with Assumption~\ref{assume_potential_model_multivalue_iv}, the proof proceeds ``as-if'' we are using a binary IV with support being $\{z, z'\}$. We now formally state the results in Corollary~\ref{corollary_local_persuasion_joint_multivalue_iv}.

\begin{corollary}\label{corollary_local_persuasion_joint_multivalue_iv}
\rm Suppose Assumption~\ref{assume_potential_model_multivalue_iv} holds, conditioning on $z, z'$ compliers (that is, $z, z' \in \supp(Z_{i})$ and $T_{i}(z) = T_{i}(z')$ does not hold almost surely), the joint distribution of potential outcome is point identified,:
\begin{align*}
\begin{split}
    \mathbb{P}[Y_{i}(1) = 1, Y_{i}(0) = 1 \mid T_{i}(z) \geq T_{i}(z')] &= \frac{\mathbb{P}[Y_{i} = 1, T_{i} = z' \mid Z_{i} = z'] - \mathbb{P}[Y_{i} = 1, T_{i} = z' | Z_{i} = z]}{\mathbb{E}[T_{i} | Z_{i} = z] - \mathbb{E}[T_{i} \mid Z_{i} = z']} \\
    \mathbb{P}[Y_{i}(1) = 1, Y_{i}(0) = 0 \mid T_{i}(z) \geq T_{i}(z')] &= \frac{\mathbb{E}[Y_{i} \mid Z_{i} = z] - \mathbb{E}[Y_{i} \mid Z_{i} = z']}{\mathbb{E}[T_{i} \mid Z_{i} = z] - \mathbb{E}[T_{i} \mid Z_{i} = z']} \\
    \mathbb{P}[Y_{i}(1) = 0, Y_{i}(0) = 0 \mid T_{i}(z) \geq T_{i}(z')] &= \frac{\mathbb{P}[Y_{i} = 0, T_{i} = z \mid Z_{i} = z] - \mathbb{P}[Y_{i} = 0, T_{i} = z | Z_{i} = z']}{\mathbb{E}[T_{i} \mid Z_{i} = z] - \mathbb{E}[T_{i} | Z_{i} = z']}.
\end{split}
\end{align*}
\end{corollary}

Just as with a discrete-valued instrument, the identification assumptions will be modified for a continuous instrument. These modifications concern the IV monotonicity and IV relevance assumptions. In this case, we use an indicator selection equation to describe the first stage selection process. With this representation, it is easy to characterize the treatment effect on different margins of self-selecting into the treatment. We also assume that at least one instrument value leads to changes in the treatment-taking behaviour. Assumption~\ref{assumption_binary_t_y_conti_z} formally states the identification assumptions for this scenario.

\begin{assumption}\label{assumption_binary_t_y_conti_z}(Binary Treatment and Outcome Model with a Continuous Instrument)

\begin{enumerate}
    \item $Y_{i}(0) \leq Y_{i}(1)$ holds almost surely, and $Y_{i}(0), Y_{i}(1) \in \{0, 1\}$,
    \item $T_{i}(z) = \mathbbm{1}\{ V_{i} \leq \nu(z) \}$, where $\nu: \mathcal{Z} \rightarrow \mathbb{R}$ is a non-trivial measurable function with respect to $z$ and assume without loss of generality that $V_{i} \sim U[0, 1]$,
    \item $Z_{i} \independent (Y_{i}(0), Y_{i}(1), V_{i}, X_{i})$.
\end{enumerate}
\end{assumption}

The identification results for identifying the joint distribution of potential outcome is presented in Corollary~\ref{corollary_joint_dist_po_conti_z}.

\begin{corollary}\label{corollary_joint_dist_po_conti_z}
\rm Assume that Assumption~\ref{assumption_binary_t_y_conti_z} holds, furthermore, assume that $\supp(P(Z_{i})) = [0, 1]$, then, the joint distribution of potential outcomes at each margin of selecting into the treatment is identified:
\begin{align*}
    &\mathbb{P}[Y_{i}(1) = 1, Y_{i}(0) = 0 \mid V_{i} = v] = \frac{\partial}{\partial v} \mathbb{E}[Y_{i} \mid P(Z_{i}) = v], \\
    &\mathbb{P}[Y_{i}(1) = Y_{i}(0) = 1 \mid V_{i} = v] \\
    &\quad = \mathbb{P}[Y_{i} = 1 \mid P(Z_{i}) = v, T_{i} = 0] - (1 - v) \frac{\partial\mathbb{P}[Y_{i} = 1 \mid P(Z_{i}) = v, T_{i} = 0]}{\partial v}, \\
    &\mathbb{P}[Y_{i}(1) = Y_{i}(0) = 0 \mid V_{i} = v] \\
    &\quad = \mathbb{P}[Y_{i} = 0 \mid P(Z_{i}) = v, T_{i} = 1] + v \frac{\partial\mathbb{P}[Y_{i} = 0 \mid P(Z_{i}) = v, T_{i} = 1]}{\partial v}.
\end{align*}
\end{corollary}

\subsection{Non-Binary Outcome}

We now discuss whether we can extend the identification of the joint distribution of potential outcomes in Lemma~\ref{corollary_local_persuasion_joint} to the case when the outcome is trinary. In the empirical study of persuasion, there are three possible outcomes: $0$ is an outside option, $1$ is the target action of persuasion, and $-1$ is any other action. Without the monotone treatment response assumption, we can classify individuals into nine types according to the potential outcomes.\footnote{\cite{jun2018identifying} does not use the conventional potential outcome notation in their discussion. \cite{jun2018identifying} first writes out the choice set facing agent $i.$ They use the following notation: $S = \{ 0, 1, -1 \}$. To write out agent $i$'s potential outcomes, \cite{jun2018identifying} uses the following notation: $Y_{i}(t) = (Y_{i0}(t), Y_{i1}(t), Y_{i,-1}(t))$, where $t \in \{0, 1\}$. $Y_{i0}(t)$ denotes whether the individual choose to take the action $0$ if the treatment is $t$. $Y_{i1}(t)$ and $Y_{i, -1}(t)$ are defined similarly. Moreover, $\sum_{j \in S} Y_{ij}(t) = 1$ for $t \in \{0, 1\}$. That is, the choices in $S$ are exclusive and exhaustive. It is easy to see that there is a duality between the notation in \cite{jun2018identifying} and conventional potential outcome notation used in Table~\ref{table_compare_po_nonbinary}.} Table~\ref{table_compare_po_nonbinary} presents the classification.

\begin{table}[ht]
\caption{Types of Individuals with Trinary Outcome}
\label{table_compare_po_nonbinary}
\begin{center}
\begin{tabular}{cccc}
$Y_{i}(0)$ & $Y_{i}(1)$ \\ \hline
$-1$         & $-1$  \\
$-1$         & $0$  \\
$-1$        & $1$ \\
$0^{**}$          & $-1^{**}$ \\
$0$          & $0$ \\
$0$          & $1$ \\
$1^{*}$      & $-1^{*}$ \\
$1^{*}$     & $0^{*}$ \\
$1$          & $1$ \\ \hline
\end{tabular}
\end{center}
\end{table}

In the previous literature, at least two types of monotone treatment response assumptions were made for trinary outcomes. \citeAppx{jun2018identifying} assume that the seventh and eighth row (those with $^{*}$) in Table~\ref{table_compare_po_nonbinary} occur with probability zero. \citeAppx{manski1997monotone} assumes that $Y_{i}(1) \geq Y_{i}(0)$ holds with probability one: the fourth row (those with $^{**}$), and the seventh and the eighth rows (those with $^{*}$) happen with zero probability.

Given the monotone treatment response assumption in \citeAppx{jun2018identifying}, we know that there are seven unknown probabilities for the joint distribution of potential outcomes among compliers. Moreover, the marginal distribution of potential outcomes among compliers is point identifiable (\citealpAppx{imbens1997estimating}). Among compliers, the joint distribution of potential outcomes is a function of the marginal distribution of potential outcomes. In other words, we have a system of linear equations with six known probabilities of the marginal distribution of potential outcomes among compliers and seven unknown probabilities of the joint distribution of potential outcomes among compliers. Therefore, the marginal distribution of potential outcomes is not point identified under the monotonicity assumption in the trinary outcome case in \citeAppx{jun2018identifying}.

A remaining question to ask is whether we can point identify the joint distribution of potential outcomes with the monotone treatment response assumption made in \citeAppx{manski1997monotone}. Again, the answer is no. The reason is that even though we have six unknowns and six equations, the information in the data is repetitive. We formally state the show the impossibility results in the following Proposition.

\begin{proposition}\label{prop_unidentifiable_joint_po_nonbinary}
\rm Assume that the potential outcomes are trinary, i.e., $Y_{i}(t) \in \{ -1, 0, 1\}$ for $t \in \{0, 1\}$. Furthermore, assume the following monotone treatment response assumption: $Y_{i}(1) \geq Y_{i}(0)$ holds with probability one. Moreover, assume assumptions 1 to 4 in Assumption~\ref{assume_potential_model} hold. Then, the joint distribution of potential outcomes among compliers is not point identified.
\end{proposition}

Even though we cannot point identify the joint distribution of potential outcomes among compliers in this case, We can still partially identify the joint distribution of potential outcomes among compliers using the approaches in \citeAppx{balke1997bounds}. For example, to construct sharp bounds for $\mathbb{P}[Y_{i}(0) = -1, Y_{i}(1) = -1 \mid T_{i}(1) > T_{i}(0)]$, we can form a linear program with the objective function being $\mathbb{P}[Y_{i}(0) = -1, Y_{i}(1) = -1 \mid T_{i}(1) > T_{i}(0)]$ and the constraints being the system of linear equations in the proof of Proposition~\ref{prop_unidentifiable_joint_po_nonbinary}.

One way to restore the point identification of the joint distribution of potential outcomes with non-binary $Y_{i}$ under the monotone treatment response and IA IV assumptions is to binarize the outcome variable. To see this, assume without loss of generality that $Y_{i}(1) \geq Y_{i}(0)$ holds almost surely. Define the following two binary random variables: $\mathbbm{1}\{ Y_{i}(1) \geq x \}$ and $\mathbbm{1}\{ Y_{i}(0) \geq x \}$ with $x \in \mathbb{R}$. Then, by the monotone treatment response, it follows immediately that $\mathbbm{1}\{ Y_{i}(1) \geq x \} \geq \mathbbm{1}\{ Y_{i}(0) \geq x \}$ holds almost surely. Thus, the results in Lemma~\ref{corollary_local_persuasion_joint} hold for the new binarized outcome variable.

\section{Profiling Compliers with a Non-Binary Instrument}\label{appx_profiling_nonbinary_z}

In Appendix~\ref{appx_identify_nonbinary_z}, we have shown that the joint distribution of potential outcomes is identifiable with a non-binary instrument. As a result, the profiling results presented in Theorem~\ref{thm_comliance_persuasion_kappa_x_t} can be readily applied to this case. The profiling results for a discrete instrument and a continuous instrument are presented in Corollary~\ref{corollary_profile_persuasion_binary_ty_discretez} and Corollary~\ref{corollary_profile_persuasion_binary_ty_contiz}, respectively. For simplicity, we only consider the case of profiling the latent types using covariates.

\begin{corollary}\label{corollary_profile_persuasion_binary_ty_discretez}
\rm Assume that Assumption~\ref{assume_potential_model_multivalue_iv} holds, and let $g: \mathbb{R} \xrightarrow[]{} \mathbb{R}$ be measurable with $\mathbb{E}[\lvert g(X_{i}) \rvert] < \infty$, then, conditioning on $z, z'$ compliers (that is, $z, z' \in \supp(Z_{i})$, $T_{i}(z) = T_{i}(z')$ does not hold almost surely, and assume without loss of generality that $T_{i}(z) \geq T_{i}(z')$ holds almost surely), the expectation of $g(X_{i})$ is identified:
\begin{align*}
    &\mathbb{E}[g(X_{i}) | Y_{i}(1) = Y_{i}(0) = 1, T_{i}(z) \geq T_{i}(z')] \\
    &= \frac{\mathbb{E}[g(X_{i}) \mathbbm{1}\{ Y_{i} = 1, T_{i} = 0 \} | Z_{i} = z'] - \mathbb{E}[g(X_{i}) \mathbbm{1}\{ Y_{i} = 1, T_{i} = 0 \} | Z_{i} = z \} ]}{\mathbb{P}[Y_{i} = 1, T_{i} = 0 | Z_{i} = z'] - \mathbb{P}[Y_{i} = 1, T_{i} = 0 | Z_{i} = z]}, \\
    &\mathbb{E}[g(X_{i}) | Y_{i}(1) = Y_{i}(0) = 0, T_{i}(z) \geq T_{i}(z')] \\
    &= \frac{\mathbb{E}[g(X_{i}) \mathbbm{1}\{ Y_{i} = 0, T_{i} = 1 \} | Z_{i} = z] - \mathbb{E}[g(X_{i}) \mathbbm{1}\{ Y_{i} = 0, T_{i} = 1 \} | Z_{i} = z']}{\mathbb{P}[Y_{i} = 0, T_{i} = 1 | Z_{i} = z] - \mathbb{P}[Y_{i} = 0, T_{i} = 1 | Z_{i} = z']}, \\
    &\mathbb{E}[g(X_{i}) | Y_{i}(1) = 1, Y_{i}(0) = 0, T_{i}(z) \geq T_{i}(z')] \\
    &= \frac{\mathbb{E}[g(X_{i}) \mathbbm{1}\{ Y_{i} = 1 \} | Z_{i} = z ] - \mathbb{E}[g(X_{i}) \mathbbm{1}\{ Y_{i} = 1 \} | Z_{i} = z' ]}{\mathbb{E}[Y_{i} | Z_{i} = z] - \mathbb{E}[Y_{i} | Z_{i} = z']}.
\end{align*}
\end{corollary}

\begin{corollary}\label{corollary_profile_persuasion_binary_ty_contiz}
\rm Assume that Assumption~\ref{assumption_binary_t_y_conti_z} holds, furthermore, assume that $\supp(P(Z_{i})) = [0, 1]$. Let $g: \mathbb{R} \xrightarrow[]{} \mathbb{R}$ be measurable with $\mathbb{E}[\lvert g(X_{i}) \rvert] < \infty$, then, conditional at each margin of selecting into the treatment, the expectation of $g(X_{i})$ is identified:
\begin{align*}
    &\mathbb{E}[g(X_{i}) \mid Y_{i}(1) = 1, Y_{i}(0) = 0, V_{i} = v] \\
    &= \frac{\frac{\partial}{\partial v} \mathbb{E}[g(X_{i}) Y_{i} \mid P(Z_{i}) = v]}{\frac{\partial}{\partial v} \mathbb{E}[Y_{i} \mid P(Z_{i}) = v]} \\
    &\mathbb{E}[g(X_{i}) \mid Y_{i}(1) = Y_{i}(0) = 1, V_{i} = v] \\
    &= \frac{\mathbb{E}[g(X_{i}) Y_{i} \mid P(Z_{i}) = v, T_{i} = 0] - (1 - v) \frac{\partial\mathbb{E}[g(X_{i}) Y_{i} \mid P(Z_{i}) = v, T_{i} = 0]}{\partial v}}{\mathbb{P}[Y_{i} = 1 \mid P(Z_{i}) = v, T_{i} = 0] - (1 - v) \frac{\partial\mathbb{P}[Y_{i} = 1 \mid P(Z_{i}) = v, T_{i} = 0]}{\partial v}} \\
    &\mathbb{E}[g(X_{i}) \mid Y_{i}(1) = Y_{i}(0) = 0, V_{i} = v] \\
    &= \frac{\mathbb{E}[g(X_{i}) \mathbbm{1}\{Y_{i} = 0\} \mid P(Z_{i}) = v, T_{i} = 1] + v \frac{\partial\mathbb{E}[g(X_{i}) \mathbbm{1}\{Y_{i} = 0\} \mid P(Z_{i}) = v, T_{i} = 1]}{\partial v}}{\mathbb{P}[Y_{i} = 0 \mid P(Z_{i}) = v, T_{i} = 1] + v \frac{\partial\mathbb{P}[Y_{i} = 0 \mid P(Z_{i}) = v, T_{i} = 1]}{\partial v}}.
\end{align*}
\end{corollary}

\section{Identification: Always-Takers and Never-Takers}\label{id_at_nt}

For always-takers, we observe their $Y_{i}(1)$. For never-takers, we observe their $Y_{i}(0)$. Therefore, the weighting method developed in Proposition~\ref{thm_kappa_g_t_x_cond_y} can be extended to always-takers and never-takers. The results are presented in Proposition~\ref{prop_local_persuasion_abadie_atnt}.

\begin{proposition}\label{prop_local_persuasion_abadie_atnt}
\rm Assume that Assume that 1 to 4 in Assumption~\ref{assume_potential_model} hold, furthermore, assume that we observe pre-treatment covariates $X_{i}$, and let $g(\cdot)$ be any measurable real function of $X_{i}$ such that $\mathbb{E}[|g(X_{i})|] < \infty$, then, for $y \in \{0, 1\}$, we have the following:
\begin{align*}
    \mathbb{E}[g(X_{i}) \mid Y_{i}(1) = y, T_{i}(1) = T_{i}(0) = 1] &= \mathbb{E}[g(X_{i}) \mid Y_{i} = y, T_{i} = 1, Z_{i} = 0] \\
    \mathbb{E}[g(X_{i}) \mid Y_{i}(0) = y, T_{i}(1) = T_{i}(0) = 0] &= \mathbb{E}[g(X_{i}) \mid Y_{i} = y, T_{i} = 0, Z_{i} = 1].
\end{align*}
\end{proposition}

With the IA IV assumption, Proposition~\ref{prop_local_persuasion_abadie_atnt} states that the conditional moments of $X_{i}$ conditioning on always-takers and their treated potential outcomes and the conditional moments of $X_{i}$ conditioning on never-takers and their untreated potential outcomes are identifiable. Furthermore, Proposition~\ref{prop_local_persuasion_abadie_atnt} implies that the conditional cumulative distribution functions are identifiable. This follows because $g(x) = \mathbbm{1}\{ X_{i} \leq x \}$ is a bounded measurable map. 

For always-takers, if we further assume the monotone treatment response, we can identify the statistical characteristic measured by pre-treatment covariates of the never-voter and always-takers. For never-takers, if we further assume the monotone treatment response, we can identify the statistical characteristic measured by pre-treatment covariates of the always-voters and never-takers.

\section{More on Estimation and Inference}\label{appx_estimation_inference}

In this appendix, we offer more detailed discussions on the estimation and inference issues related to the estimands proposed in Section 4 and 5. Our first focus is on the estimation and inference results with strong identification. Afterward, we shift our discussion to the inference results when identification is weak.

\subsection{Estimation and Inference under Strong Identification}

Recall that our identification results give us the following $\beta_{IV}$ estimand:
\begin{align*}
    \beta_{IV} = \frac{\mathbb{E}[f(X_{i}, Y_{i}, T_{i}) \mid Z_{i} = 1] - \mathbb{E}[f(X_{i}, Y_{i}, T_{i}) \mid Z_{i} = 0]}{\mathbb{E}[h(Y_{i}, T_{i}) \mid Z_{i} = 1] - \mathbb{E}[h(Y_{i}, T_{i}) \mid Z_{i} = 0]}.
\end{align*}
We can use the sample analog to estimate $\beta_{IV}$:
\begin{align*}
    \hat{\boldsymbol{\beta}}_{IV} = \left( \frac{1}{n} \sum_{i = 1}^{n} \begin{pmatrix}
           1 \\
           Z_{i}
         \end{pmatrix} (1, h(Y_{i}, T_{i})) \right)^{-1} \left( \frac{1}{n} \sum_{i = 1}^{n} \begin{pmatrix}
           1 \\
           Z_{i}
         \end{pmatrix} f(X_{i}, Y_{i}, T_{i}) \right),
\end{align*}
with $\hat{\beta}_{IV}$ being the second component of $\hat{\boldsymbol{\beta}}_{IV}$. Using a standard argument (e.g., see Chapter 12 in \citeAppx{hansen2022econometrics}), we can show the consistency and asymptotic normality of $\hat{\boldsymbol{\beta}}_{IV}$ under suitable regularity conditions. We now formally claim the results below.

\begin{proposition}
\rm Assume that the following conditions hold:
\begin{enumerate}
    \item $\mathbb{E}[f(X_{i}, Y_{i}, T_{i})^{4}] < \infty$,
    \item $\mathbb{E}\left[\begin{pmatrix}
           1 \\
           Z_{i}
    \end{pmatrix} (1, Z_{i})\right]$ is positive definite,
    \item $\mathbb{E}\left[\begin{pmatrix}
           1 \\
           Z_{i}
    \end{pmatrix} (1, h(Y_{i}, T_{i}))\right]$ is rull rank,
    \item $\mathbb{E}\left[\begin{pmatrix}
           1 \\
           Z_{i}
    \end{pmatrix} e_{i} \right] = 0$, where $e_{i}$ is the residual from regressing $f(X_{i}, Y_{i}, T_{i})$ on $h(Y_{i}, T_{i})$,
    \item $\mathbb{E}[h(Y_{i}, T_{i})^{4}] < \infty$,
    \item $\mathbb{E}[Z_{i}^{4}] < \infty$,
    \item $\Omega = \mathbb{E}\left[\begin{pmatrix}
           1 \\
           Z_{i}
    \end{pmatrix} (1, Z_{i}) e_{i} \right]$ is positive definite,
\end{enumerate}
then, $\sqrt{n} \left( \hat{\boldsymbol{\beta}}_{IV} - \boldsymbol{\beta}_{IV} \right)$ is asymptotically normal:
\begin{align*}
    \sqrt{n} \left( \hat{\boldsymbol{\beta}}_{IV} - \boldsymbol{\beta}_{IV} \right) \xrightarrow[]{\mathcal{D}} N\left(0, \mathbb{E}\left[\begin{pmatrix}
           1 \\
           Z_{i}
    \end{pmatrix} (1, h(X_{i}, T_{i})) \right]^{-1} \Omega \mathbb{E}\left[\begin{pmatrix}
           1 \\
           h(X_{i}, T_{i})
    \end{pmatrix} (1, Z_{i}) \right]^{-1} \right).
\end{align*}
Moreover, a consistent estimator for $\mathbb{E}\left[\begin{pmatrix}
           1 \\
           Z_{i}
    \end{pmatrix} (1, h(X_{i}, T_{i})) \right]^{-1} \Omega \mathbb{E}\left[\begin{pmatrix}
           1 \\
           h(X_{i}, T_{i})
\end{pmatrix} (1, Z_{i}) \right]^{-1}$ is:
\begin{align*}
    \left( \frac{1}{n} \sum_{i = 1}^{n} \begin{pmatrix}
           1 \\
           Z_{i}
    \end{pmatrix} (1, h(X_{i}, T_{i})) \right)^{-1} \hat{\Omega} \left( \frac{1}{n} \sum_{i = 1}^{n} \begin{pmatrix}
           1 \\
           h(X_{i}, T_{i})
\end{pmatrix} (1, Z_{i}) \right)^{-1},
\end{align*}
where $\hat{\Omega} = \left( \frac{1}{n} \sum_{i = 1}^{n} \begin{pmatrix}
           1 \\
           Z_{i}
    \end{pmatrix} (1, Z_{i}) \left( f(X_{i}, Y_{i}, T_{i}) - (1, h(Y_{i}, T_{i})) \hat{\boldsymbol{\beta}}_{IV} \right) \right)$.
\end{proposition}

Before we proceed, we now give a remark on the consistency of the estimator we proposed. For the sake of simplicity, assume that $X_{i} \in \mathbb{R}$. Let $g(X_{i}) = \mathbbm{1}\{ X_{i} \leq x \}$, Theorem~\ref{thm_kappa_g_t_x_cond_y} shows that we can point identify the conditional distribution function conditional on $[Y_{i}(0) = 0, T_{i}(1) > T_{i}(0)]$:
\begin{align*}
    &\mathbb{P}[ X_{i} \leq x \mid Y_{i}(0) = 0, T_{i}(1) > T_{i}(0) ] \\
	&= \frac{\mathbb{P}[ X_{i}  \leq x, Y_{i} = 0, T_{i} = 0 \mid Z_{i} = 0] - \mathbb{P}[ X_{i}  \leq x, Y_{i} = 0, T_{i} = 0 \mid Z_{i} = 1]}{\mathbb{P}[Y_{i} = 0, T_{i} = 0 \mid Z_{i} = 0] - \mathbb{P}[Y_{i} = 0, T_{i} = 0 \mid Z_{i} = 1]}.
\end{align*}
It is easy to see that $\hat{\mathbb{P}}[ X_{i} \leq x \mid Y_{i}(0) = 0, T_{i}(1) > T_{i}(0) ]$ is a (pointwise) consistent estimator for $\mathbb{P}[ X_{i} \leq x \mid Y_{i}(0) = 0, T_{i}(1) > T_{i}(0) ]$. By the same idea in the Glivenko-Cantelli Theorem (see, e.g., Theorem 2.4.7 in \citeAppx{durrett2010probability}), we can strengthen the pointwise consistency to uniform consistency:
\begin{align*}
    \sup_{x \in \mathbb{R}} \left| \hat{\mathbb{P}}[ X_{i} \leq x \mid Y_{i}(0) = 0, T_{i}(1) > T_{i}(0) ] - \mathbb{P}[ X_{i} \leq x \mid Y_{i}(0) = 0, T_{i}(1) > T_{i}(0) ] \right| \xrightarrow[]{\mathbb{P}} 0.
\end{align*}
We prove this uniform consistent result in Appendix~\ref{appx_gelivenko_cantelli}.

\subsection{An Anderson-Rubin Test under Weak Identification}

Note that the estimand in Equation~\ref{equation_wald_estimand} is a function of two regression coefficients:
\begin{align*}
    p = \frac{\beta_{1}}{\beta_{2}} \equiv \frac{\mathbb{E}[f(X_{i}, Y_{i}, T_{i}) \mid Z_{i} = 1] - \mathbb{E}[f(X_{i}, Y_{i}, T_{i}) \mid Z_{i} = 0]}{\mathbb{E}[h(Y_{i}, T_{i}) \mid Z_{i} = 1] - \mathbb{E}[h(Y_{i}, T_{i}) \mid Z_{i} = 0]}.
\end{align*}
A concern regarding the asymptotic approximation discussed in the previous section is that the denominator $\beta_{2}$ may be close to zero. When faced with weak identification, the asymptotic approximation discussed earlier may not perform well. Fortunately, in the current exact identified scenario, we can use the Anderson-Rubin test to circumvent the issue of weak identification.

Note that under the null hypothesis $H_{0}: p = p_{0}$, we have that $p_{0} \beta_{2} - \beta_{1} = 0$. Therefore, by using the delta method, the limiting distribution of $\sqrt{n} (p_{0} \hat{\beta}_{1} - \hat{\beta}_{2})$ under $H_{0}$ is:
\begin{align*}
    \sqrt{n} (p_{0} \hat{\beta}_{1} - \hat{\beta}_{2}) \xrightarrow[]{\mathcal{D}} N(0, \gamma),
\end{align*}
where $\gamma = \var(\beta_{1}) - 2 p_{0} \cov(\beta_{1}, \beta_{2})) + p_{0}^{2} \var(\beta_{2})$.

Therefore, a test statistic is:
\begin{align*}
    T_{n} = \frac{n (p_{0}\hat{\beta}_{1} - \hat{\beta_{2}})^{2}}{\hat{\gamma}},
\end{align*}
where $\hat{\gamma}$ is a consistent estimator for $\gamma$. By Slutsky's Lemma, we further know that:
\begin{align*}
    T_{n} \xrightarrow[]{\mathcal{D}} \chi(1).
\end{align*}

Using the AR statistic, we can form an AR test of $H_{0}: p = p_{0}$ as:
\begin{align*}
    \phi_{AR}(p_{0}) = \mathbbm{1}\{ T_{n} > \chi^{2}_{1, 1 - \alpha}\},
\end{align*}
where $\chi^{2}_{1, 1 - \alpha}$ is the $1 - \alpha$ quantile of $\chi^{2}_{1}$ distribution. As noted by \citeAppx{staiger1997instrumental}, this yields a size-$\alpha$ test that is robust to weak identification. We then can form a level $1 - \alpha$ weak-identification-robust confidence set by collecting the nonrejected values.

\section{Implementing the Test for Identification Assumptions}\label{appx_implement_sharp_test_id_assumption}

Recall that in Section~\ref{sharp_test_id_assumption}, the test statistic is given by:
\begin{align*}
    T_{n} \equiv \inf_{\mathbf{p} \geq \mathbf{0}: B \mathbf{p} = 1} \sqrt{n} \left| A_{\text{obs}} \mathbf{p} - \hat{\mathbf{b}} \right|.
\end{align*}
To compute the test statistic, we choose the $\ell_{2}$ norm. Thus, the minimizer to the minimization problem in the test statistic can be obtained by solving:
\begin{align*}
    &\min_{\mathbf{p}} \left| \left| A_{\text{obs}} \mathbf{p} - \hat{\mathbf{b}} \right| \right|_{2} \\
    &\text{subject to } \mathbf{p} \geq \mathbf{0}, \sum_{i = 1}^{\dim(\mathbf{p})} p_{i} = 1,
\end{align*}
where the inequality in the constraint is interpreted to hold component-wise. Note that the minimizer of the optimization problem above is equivalent to the minimizer of the following minimization problem:
\begin{align*}
    &\min_{\mathbf{p}} \mathbf{p}^{T} A_{\text{obs}}^{T} A_{\text{obs}} \mathbf{p} - 2 \mathbf{p}^{T} A_{\text{obs}}^{T} \hat{\mathbf{b}} \\
    &\text{subject to } \mathbf{p} \geq \mathbf{0}, \sum_{i = 1}^{\dim(\mathbf{p})} p_{i} = 1,
\end{align*}
The minimization problem above is a convex problem (\citealpAppx{boyd2004convex}), and can be efficiently solved by using \textbf{CVXR} package in \texttt{R} (\citealpAppx{fu2017cvxr}). 

After solving the optimal $\mathbf{p}^{*}$, we then can compute the test statistics by computing:
\begin{align*}
    T_{n} = \sqrt{n} \left\vert A_{\text{obs}} \mathbf{p}^{*} - \hat{\mathbf{b}} \right\vert.
\end{align*}

\bibliographystyleAppx{chicago}
\bibliographyAppx{reference_appx}

\end{appendices}

\end{document}